 \documentclass[11pt]{article}

\usepackage{amsmath,amsfonts,amssymb}
\usepackage{doi}

 \usepackage[numbers,sort&compress]{natbib}
\usepackage{graphicx}

\usepackage{hyperref}
\hypersetup{
   colorlinks   =  true,
    linkcolor    = blue,
    citecolor    = red,
     urlcolor	=blue,
}

\newcommand{\sect}[1]{\setcounter{equation}{0}\section{#1}}
\newcommand{\subsect}[1]{\subsection{#1}}

\renewcommand{\theequation}{\arabic{section}.\arabic{equation}}
\newtheorem{proposition}{Proposition}
\newtheorem{conjecture}{Conjecture}

\newcommand\be{\begin{equation}}
\newcommand\ee{\end{equation}}
\newcommand\bea{\begin{eqnarray}}
\newcommand\eea{\end{eqnarray}}

\newcommand{\cH}{\mathcal{H}} %Caligraphic H
\newcommand{\cC}{\mathcal{C}} %Caligraphic C
\newcommand{\cI}{\mathcal{I}} %Caligraphic I
\newcommand{\hcH}{ \hat{\mathcal{H}} } %hat  Caligraphic H
\newcommand{\hcC}{\hat{\mathcal{C}}} %hat Caligraphic C
\newcommand{\cK}{\mathcal{K}} %Caligraphic K
\newcommand{\hcK}{\hat {\mathcal{K}}} %hat Caligraphic K

  \newcommand{\hP}{\hat {\mathcal{P}}} %hat polynomial P

\newcommand{\hI}{{\hat I}} %hat I
\newcommand{\hq}{{\hat q}} %hat q  
\newcommand{\hp}{{\hat p}} %hat p  

\newcommand{\B}{{B}} % number operator B

\newcommand{\hB}{{\hat B}} %hat B
\newcommand{\hbm}{{\hat b}} %hat b
\newcommand{\hbp}{{\hat b}^\dagger} %hat b dagger

 \newcommand{\ri}{{\rm i}} %hat i complex
 \newcommand{\kk}{k} %  index
  \newcommand{\mmu}{\mu} %  mu
  \newcommand{\nnu}{\nu} %  nu

\newcommand{\n}{ {n}} %  principal quantum number
\newcommand{\fn}{\mathfrak{n}} % mathfrak principal quantum number

 \DeclareMathOperator\spn{span}
  \newcommand{\otra}{\lambda}
  
  \def\>#1{{\mathbf #1}} %bold

\newcommand{\bhq}{ {\hat{\mathbf{q}} }} %hat bold q  
\newcommand{\bhp} { {\hat{\mathbf{p}} }}  %hat bold p  

\begin{document}

 \noindent
 {\Large \bf
 Generalized quantum Zernike Hamiltonians: Polynomial\\[5pt]Higgs-type algebras   and algebraic derivation of the spectrum}

\medskip

\medskip

\noindent
{\sc  Rutwig Campoamor-Stursberg$^{1,2,\star}$, Francisco J.~Herranz$^{3,*}$, Danilo Latini$^{4,5,\dagger}$, \\[2pt]
 Ian Marquette$^{6,\ddagger}$ and Alfonso Blasco$^{3,\sharp}$
}

\medskip

 \noindent
$^1$ Instituto de Matem\'atica Interdisciplinar, Universidad Complutense de Madrid, E-28040 Madrid,  Spain

\noindent
$^2$ Departamento de \'Algebra, Geometr\'{\i}a y Topolog\'{\i}a,  Facultad de Ciencias
Matem\'aticas, Universidad Complutense de Madrid, Plaza de Ciencias 3, E-28040 Madrid, Spain

\noindent
{$^3$ Departamento de F\'isica, Universidad de Burgos,
E-09001 Burgos, Spain}

\noindent
{$^4$ Dipartimento di Matematica ``Federigo Enriques", Universit\`a degli Studi di Milano, Via C. Saldini 50, 20133 Milano, Italy}

\noindent
{$^5$  INFN Sezione di Milano, Via G. Celoria 16, 20133 Milano, Italy}

\noindent
$^6$ Department of Mathematical and Physical Sciences, La Trobe University, Bendigo, VIC 3552, Australia

 \medskip

\noindent   
{ {\footnotesize
 {$^\star$\href{mailto:rutwig@ucm.es}{\texttt{rutwig@ucm.es}}\ \  $^*$\href{mailto:fjherranz@ubu.es}{\texttt{fjherranz@ubu.es}}\ \   $^\dagger$\href{mailto:danilo.latini@unimi.it}{\texttt{danilo.latini@unimi.it}}\ \   $^\ddagger$\href{mailto:i.marquette@latrobe.edu.au}{\texttt{i.marquette@latrobe.edu.au}}\ \   $^\sharp$\href{mailto:ablasco@ubu.es}{\texttt{ablasco@ubu.es}}
 }
 }

\begin{abstract}
\noindent
We consider the quantum analog of the generalized Zernike systems given by the Hamiltonian:
$$\hcH_N =\hat{p}_1^2+\hat{p}_2^2+\sum_{k=1}^N \gamma_k (\hat{q}_1 \hat{p}_1+\hat{q}_2 \hat{p}_2)^k ,$$
with canonical operators $\hat{q}_i,\, \hat{p}_i$ and arbitrary coefficients $\gamma_k$.
This two-dimensional quantum model, besides the conservation of the angular momentum, exhibits higher-order integrals of motion within the enveloping algebra of the Heisenberg algebra   in two dimensions. By constructing suitable combinations of these integrals, we uncover a polynomial Higgs-type symmetry algebra that, through an appropriate change of basis, gives rise to a deformed oscillator algebra. The associated structure function $\Phi$ is shown to factorize into two commuting components $\Phi=\Phi_1 \Phi_2$. This framework enables an algebraic determination of the possible energy spectra of the model for the cases $1\le N \le 5$,  the case $N=1$ being canonically equivalent to the harmonic oscillator. Based on these findings, we propose two conjectures which generalize the results for all  $N\ge 1$ and any value of the coefficients $\gamma_k$. In addition, all of these results can be interpreted as  higher-order superintegrable perturbations of the original quantum Zernike system corresponding to $N=2$, which are also analyzed and  applied to the isotropic oscillator on the sphere, hyperbolic and Euclidean spaces.

\end{abstract}
\medskip

\noindent 
{\bf To the memory of our colleague and friend 
Prof. George Samvelovich Pogosyan}

\bigskip

 
\medskip



  \newpage

\tableofcontents

%%%%%%%%%%%%%%%%%%%%%%%%%%%%%%%%%%%%%%%%%%%%%%

\sect{Introduction}

The so-called Zernike system, that emerged in the context of wavefront descriptions in optics~\cite{Zernike1934}, has turned out to be an extremely interesting model, with many physically and mathematically relevant applications. Besides being the first case in a hierarchy of (classically) superintegrable systems~\cite{zk2023}, its quantum version determines a new class of orthogonal polynomials, the Zernike polynomials, of great interest in the treatment of optical aberrations. In addition, it has been shown that the quantum symmetries of the system give rise to a cubic Higgs algebra  $\mathfrak{sl}^{(3)}(2,\mathbb{R})$. The latter fact, in combination with recent work on polynomial algebras in connection with the superintegrability property (see e.g.~\cite{RCM} and references therein), suggest to look for a quantization of the generalized classical Zernike Hamiltonians considered in~\cite{zk2023} in such a way that the corresponding quantum symmetry algebra is a higher-order polynomial  Higgs-type algebra.

The two-dimensional Zernike system is defined by means of the following quantum Hamiltonian $ \hcH_{\rm Zk}$  with eigenvalue equation (with normalized units $\hbar=1$)  
 \be
\hcH_{\rm Zk}  \psi( \mathbf{q} )  =\left(  -  \nabla^2  -\beta\,   \>q \boldsymbol{\cdot} \nabla-   \alpha  
\bigl(\>q \boldsymbol{\cdot} \nabla \bigr)^2 \right) \psi( \mathbf{q} ) =E\,\psi( \mathbf{q} ) ,
\label{a1}
\ee
   where  we have denoted
\be
\>q=(q_1,q_2), \qquad \nabla=\left(\frac{\partial}{\partial q_1},\frac{\partial}{\partial q_2}\right),\qquad 
\>q \boldsymbol{\cdot} \nabla = q_1\frac{\partial}{\partial q_1 }+q_2\frac{\partial}{\partial q_2},
\nonumber
\ee
such that $\alpha$ and $\beta$ are two real parameters. It turns out that  the Zernike system exhibits interesting structural features of great relevance in quantum optics, which have motivated its detailed study in recent years (see~\cite{PSWY2017,Atakishiyev2017,PWY2017a,Fordy2018,Atakishiyev2019,Wolf2020}  and references therein). Among these important properties, the Zernike system is superintegrable, that is, it admits two algebraically independent quantum observables that commute with the Hamiltonian $ \hcH_{\rm Zk}$. Therefore, the corresponding spectrum presents a maximal degeneracy.  We also recall that, as shown in~\cite{Zernike1934},  among the possible values for the real parameters $\alpha$ and $\beta$, the case with $\alpha=-1$ and $\beta=-2$ is the only one that provides a self-adjoint  operator $ \hcH_{\rm Zk}$  under the inner product over the unit disk with $|\>q|\le 1$, which    has been fully studied   by Pogosyan {\em et al} in~\cite{PSWY2017}, leading to the Zernike
polynomials (see e.g.~\cite{Wunsche2005,Zpolynomials1,Zpolynomials2} and references therein).

   On the other hand, the classical  Hamiltonian counterpart of $ \hcH_{\rm Zk}$ (\ref{a1}), also considered by Pogosyan {\em et al} in~\cite{PWY2017zernike}, adopts the following form
   \begin{equation}
\cH_{\rm Zk} =  p_1^2 + p_2^2 -\ri \beta (q_1p_1+ q_2 p_2  )+\alpha (q_1p_1+ q_2 p_2  )^2,
\label{a3}
\end{equation}
where $q_i$ and $p_i$ are generic canonical variables fulfilling the Poisson bracket $\{q_i,p_j\}=\delta_{ij}$. The Hamiltonian $\cH_{\rm Zk}$ determines a classical superintegrable system  (in the Liouville sense~\cite{Perelomov}), which means that it admits two constants of the motion, in this case  quadratic in the momenta.  Explicitly, these are the usual 
  angular momentum
\be
\cC = q_1 p_2 - q_2 p_1
\label{a5}
\ee
together with a function given by
\be
\mathcal I=  p_2^2   -\ri \beta  q_2p_2   +\alpha  \bigl( q_1^2 + q_2^2 \bigr)p_2^2 .
\label{a5b}
\ee
Thus, the Hamiltonian $\cH_{\rm Zk}$ Poisson-commutes with $\cC$ and $ \mathcal I$, although   these constants  are not in involution:
 $$
\big\{\cH_{\rm Zk} , \cC \big \}=\big\{\cH_{\rm Zk} ,  \mathcal I\} =0,\qquad \big\{ \cC ,  \mathcal I\} \ne 0.
$$
In this respect, we recall that the existence of {\em one} constant of the motion implies that $\cH_{\rm Zk}$ determines a classical  {\em  integrable} system. Moreover, as the three functions  $\cH_{\rm Zk}$, $\cC$ and $ \mathcal I$ are 
functionally independent, the classical Zernike Hamiltonian is  {\em  superintegrable}, and  therefore endowed with the maximum number of constants of the motion for this dimension.
As a consequence, all bounded trajectories are closed and periodic, which  for $\cH_{\rm Zk}$   are given by  ellipses~\cite{PWY2017zernike}.

 In addition,  another constant of the motion can be obtained immediately from   $\mathcal I$ (\ref{a5b}) in the form
 \be
\mathcal I'=  p_1^2   -\ri \beta  q_1p_1   +\alpha  \bigl( q_1^2 + q_2^2 \bigr)p_1^2 ,\qquad \big\{\cH_{\rm Zk} ,  \mathcal I'\} =0, 
\label{a5c}
\ee
which   cannot be functionally independent with respect to the previous ones, thereby verifying
  the following relation   
$$
\cH_{\rm Zk} = \mathcal I+\mathcal I' - \alpha \cC^2 .
$$
The introduction of $\mathcal I'$ enables  the construction of the symmetry algebra of $\cH_{\rm Zk}$, understood as  the algebra closed by its constants of the motion. According to~\cite{PWY2017zernike}, if we define 
 \begin{equation}
   \begin{split}
   \mathcal L_1&:=\frac 12\mathcal C , \qquad  \mathcal L_2:=\frac 12\bigl(\mathcal I'  -  \mathcal I \bigr)  , \\[2pt]
 \mathcal L_3&:= \{  \mathcal L_1,\mathcal L_2\}= \left( 1+ \alpha \bigl(q_1^2+q_2^2\bigr)  \right)p_1 p_2 - \frac 12 \ri \beta (q_1 p_2 + q_2 p_1) ,
 \end{split}
\label{a5e}
\end{equation}
we obtain the following Poisson brackets
\be
\{ \mathcal L_1,\mathcal  L_2\}=\mathcal L_3, \qquad \{ \mathcal L_1,\mathcal  L_3\}=-\mathcal L_2, \qquad  \{ \mathcal L_2, \mathcal L_3\}=\big( \beta^2 - 2\alpha \mathcal H_{\rm Zk} \big)\mathcal L_1 -8  \alpha^2  \mathcal L_1^3  .
\label{za5fd}
\ee
This structure determines a cubic polynomial Poisson algebra, 
 which for the values $\alpha=-1$ and $\beta=-2$~\cite{PWY2017zernike} reproduces     the so-called Higgs algebra, {\em i.e.}, the symmetry algebra of   the 
      Higgs  oscillator on the  sphere~\cite{Higgs}. In fact, we recall that if different values of the parameter $\alpha$  are allowed,  the classical Zernike Hamiltonian $\cH_{\rm Zk}$ (\ref{a3}) has been alternatively interpreted in~\cite{zk2023} as the two-dimensional isotropic oscillator on the  sphere, hyperbolic and Euclidean spaces  for $\alpha<0$, $\alpha>0$ and $\alpha=0$, respectively. Therefore, from this point of view, the parameter
$\alpha$ determines the constant curvature of the underlying space, while  
  $\beta$ is related to the frequency of the corresponding oscillator.
Observe that  for the  Euclidean oscillator with $\alpha=0$, the   polynomial algebra (\ref{za5fd})
reduces to  the Lie algebra $\mathfrak{so} (3)$.
It should also be noted that,   very recently,  it has been shown in~\cite{ZernikeRevisited2026}  that the `imaginary $\beta$-term' in $\cH_{\rm Zk}$ (\ref{a3})
 is provided by  a vector potential that determines a magnetic field,  which is a pure gauge and can be eliminated by a  canonical transformation for any value of $\alpha$.

A first superintegrable generalization of the classical Zernike Hamiltonian (\ref{a3}) was considered and analyzed in~\cite{zk2023} in the form 
 \begin{equation}
\cH_N =  p_1^2 + p_2^2 +\sum_{\kk=1}^N \gamma_\kk(q_1 p_1 + q_2 p_2)^\kk ,
\label{a4}
\end{equation}
where   $\gamma_k$ are arbitrary coefficients that can be either pure imaginary or real numbers. Two constants of the motion for the Hamiltonian $\cH_N$ are the  angular momentum (\ref{a5}) a certain function $\cI_N$, which is of $N^{{\rm th}}$-order in the momenta and has been  obtained explicitly in~\cite{zk2023} for arbitrary index $N$. The classical Zernike system $\cH_{\rm Zk}$ is recovered from $\cH_N$ (\ref{a4}) for the quadratic case with $N=2$ by setting
\be
\gamma_1=-\rm i \beta,\qquad \gamma_2=\alpha ,\qquad \alpha,\beta\in\mathbb R.
\label{a6}
\ee
In this context, $\cH_N$ can be interpreted as a superintegrable perturbation of $\cH_{\rm Zk} $ for $N\ge 3$, and such that the bounded trajectories, albeit preserving the property of being closed and periodic,  are no longer ellipses, but some `deformation' of them that increases with the values of $\gamma_k$ ($k\ge 3)$, as shown in~\cite{zk2023}. More recently, an additional generalization of the Hamiltonian \eqref{a4} has been proposed in~\cite{Gonera2024} in terms of generic analytical functions $F\left(q_1 p_1 + q_2 p_2\right)$, showing that the superintegrability property actually arises from the particular structure of the Hamiltonian.

The aim of this work is twofold. In the first place, we propose a generalized quantum Zernike Hamiltonian  $\hcH_N $  by imposing that the following requirements are satisfied:\newline 
(i) The superintegrability property is preserved for any $N$.
\newline
(ii) For $N=2$, the quantum Zernike system  $\hcH_{\rm Zk}$ in (\ref{a1}) is recovered.
\newline
(iii) The classical counterpart of $\hcH_N $ is given by $\cH_N$ in (\ref{a4}).

As second objective, we derive the corresponding spectrum of $\hcH_{\rm Zk}$ by means of an algebraic procedure that involves the quantum symmetries of $\hcH_N $, {\em i.e.}, algebraically independent quantum observables commuting with $\hcH_N $, which are shown to be of $N^{{\rm th}}$-order in the momenta.  

 Furthermore, similarly to what occurs with $\cH_{\rm Zk}\equiv \cH_2$,  the quantum Zernike system $\hcH_{\rm Zk}\equiv \hcH_2$ can be interpreted   as a quantum oscillator on the sphere, hyperbolic and Euclidean spaces; the latter actually corresponds to the case with $N=1$   ($\gamma_2=\alpha=0$).  As a direct consequence of this result, our findings provide superintegrable perturbations of such (curved) quantum oscillators for $N\ge 3$, which we also study in detail.

   The structure of the paper is as follows: In section \ref{s2} we present the algebraic procedure used to obtain the quantum symmetries and spectrum for the generalized quantum Zernike Hamiltonian $\hcH_N$ in (\ref{a4}). In section~\ref{s3}, our approach is first applied to the proper (quadratic) quantum Zernike Hamiltonian $ \hcH_{\rm Zk}$ (\ref{a1}), recovering  known results. As a byproduct, the associated  quantum curved oscillators are also deduced and studied.    Section~\ref{s4}  is devoted to the cubic  system, with $N=3$, which is  explicitly solved.  In addition, we stress that the cubic case is shown to introduce a superintegrable perturbation of the quantum isotropic oscillator on the above three Riemannian spaces that are fully analyzed in section~\ref{s42}. 
  In particular, two types of perturbations arise,  called  spherical or hyperbolic cubic perturbations, which hold for each of the quantum oscillators in the sphere, hyperbolic and Euclidean spaces.  
 Therefore, in addition to the two curved oscillators, which cover the Higgs oscillator, this application also yields two new Euclidean quantum oscillators.

Due to ordering problems in the quantization, the computational difficulties increase exponentially for higher values of $N$.  To facilitate the reading of the manuscript, the fourth-order system with $N=4$  is addressed in Appendix~\ref{appendixA}, where the construction is developed along the same lines as for the previous cubic system and the corresponding results are presented in detail. The main results for the fifth-order system with $N=5$ are summarized in Appendix~\ref{appendixB}.

 It should be observed that  the general expression for arbitrary $N$ remains as an open problem; in fact, even in the classical framework~\cite{zk2023}, the general construction is quite involved. Nevertheless,   taking into account the results up to $N=5$,  in section~\ref{s5} we present  two conjectures for the spectrum of $\hcH_N$    (\ref{a4}) with arbitrary $N$. The paper concludes with some remarks and some open problems whose detailed analysis will be considered in the future.

  %%%%%%%%%%%%%%%%%%%%%%%%%%%%%%%%%%%%%%%%%%%%%%
 
 \sect{Generalized quantum Zernike Hamiltonians}
 \label{s2}

Let us consider  the usual quantum position $\bhq=(\hq_1,\hq_2)$ and momenta  $\bhp=(\hp_1,\hp_2)$ operators, with canonical Lie brackets and differential realization given by
\be
[\hat q_i,\hat q_j]=[\hat p_i,\hat p_j]=0,\qquad [\hat q_i,\hat p_j]={\rm i} \,{\rm Id}\, \delta_{ij},\qquad \hat q_i \psi( \mathbf{q} ) =q_i\psi( \mathbf{q} ),\qquad \hat p_i \psi( \mathbf{q} )=-{\rm i} \,\frac{\partial   \psi( \mathbf{q} )}{\partial q_i} , 
\label{b1}
\ee
where  ${\rm Id}$ denotes the identity operator and, for clarity in the exposition, we normalize units by setting $\hbar=1$.
Under this realization, the quantum Zernike Hamiltonian (\ref{a1}) is written as 
   \begin{equation}
   \begin{split}
\hcH_{\rm Zk} &=  \hp_1^2 + \hp_2^2 -\ri \beta (\hq_1\hp_1+ \hq_2 \hp_2  )+\alpha (\hq_1\hp_1+ \hq_2 \hp_2  )^2 \\[2pt]
&=\bhp^2 -\ri \beta (\bhq\boldsymbol{\cdot} \bhp  )+\alpha (\bhq\boldsymbol{\cdot} \bhp  )^2.
\end{split}
\label{b2}
\end{equation}
Taking into account the latter expression and the classical superintegrable Hamiltonian $\cH_N$ (\ref{a4}) 
we shall consider, as a suitable quantum analogue, the following quantum Hamiltonian~\cite{zk2023} 
\be
\hcH_N  =  \bhp^2 + \sum_{\kk=1}^N \gamma_\kk ( \bhq  \boldsymbol{\cdot} \bhp)^\kk   ,
\label{b3}
\ee
where  $\gamma_k$ are arbitrary (real or pure imaginary) coefficients. This
leads to the Schr\"odinger equation given by
 \be
    \begin{split}
\hcH_N \psi( \mathbf{q} )  &=\left(  - \nabla^2+ \sum_{\kk=1}^N \gamma_\kk(-{\rm i}  )^\kk
\bigl(\>q \boldsymbol{\cdot} \nabla \bigr)^\kk \right)\psi( \mathbf{q} )=E\,\psi( \mathbf{q} )\\[4pt]
&=\left(  - \left(\frac{\partial^2  }{\partial q_1^2}+\frac{\partial^2  }{\partial q_2^2} \right)+ \sum_{\kk=1}^N \gamma_\kk(-{\rm i}  )^\kk
\left( q_1 \frac{\partial  }{\partial q_1}+   q_2 \frac{\partial  }{\partial q_2~} \right)^{\!\kk} \right)\psi( \mathbf{q} ) =E\,\psi( \mathbf{q} ) ,
   \end{split}
\label{b4} 
\ee
which generalizes to arbitrary values of $N\ge 3$ the Hamiltonian $\hcH_{\rm Zk}$ (\ref{a1}), that is recovered with the choices (\ref{a6}). It is important to observe that, due to the term ordering problems arising in the quantization of $\cH_N$ (\ref{a4}), as well as for the corresponding constants of the motion, there are many possibilities for the quantum construction. Nevertheless, we will show here that the Hamiltonian $\hcH_N $ in (\ref{b3}) determines a quantum superintegrable system by explicitly finding its quantum symmetries and then computing its spectrum for arbitrary coefficients $\gamma_\kk$. As expected, for increasing values of $N$, some of the  computations become extremely cumbersome, requiring the use of symbolic computer packages, such as the {\em Wolfram Mathematica\textsuperscript{\textregistered}} software system.

The first step in our construction is to obtain the quantum symmetries of $\hcH_N$ ensuring the superintegrability property.
One of these symmetries  corresponds to the quantum angular momentum operator 
\be
\hcC = \hq_1 \hp_2 - \hq_2 \hp_1,
\label{b5}
\ee
as it commutes with $ \bhq  \boldsymbol{\cdot}\bhp$. This shows that $\hcH_N$ is a quantum integrable Hamiltonian. Due to the structure of the Hamiltonian, we expect the additional quantum symmetry for $\hcH_N$ to be of higher-order in the momenta. Following this assumption, we consider the enveloping algebra of the Heisenberg algebra in two dimensions spanned by  $\bhq$ and  $\bhp$ (\ref{b1}), to which the identity operator ${\rm Id}$ is added as fifth generator, and introduce the following grading 
\begin{equation}
 |\hq_1|= |\hq_2|=1,\qquad |\hp_1|=  |\hp_2|=-1 .
\nonumber
\end{equation} 
Next,  we look for monomials $\hP_\ell^{(s)}(\bhq,\bhp)$ of degree $\ell\leq 2N$ such that the total grading of the monomial is equal to 0. The grading is assigned via the prescription 
\begin{equation}
 | \hq_1^a \, \hq_2^b \, \hp_1^c \, \hp_2^d |= a+b-c-d ,
 \nonumber
\end{equation} 
  where $a$, $b$, $c$ and $d$ are integers.
In this way, we construct the following operators of $N^{{\rm th}}$-order in the momenta
\begin{equation}
\hI_N = \hp_2^2 + \sum_{ {  \ell=1}}^{2N} \sum_{{  s=1}}^{{\rm dim} \hP_{\! \ell} }A_\ell^{(s)} \hP_\ell^{(s)} ,\qquad  \hI_N' = \hp_1^2 + \sum_{{  \ell=1}}^{2N} \sum_{{  s=1}}^{{\rm dim} \hP_{\! \ell} }B_\ell^{(s)} \hP_\ell^{(s)} ,
  \label{b6b}
\end{equation}
  where $A_\ell^{(s)}$ and $B_\ell^{(s)}$ are certain coefficients to be determined. These operators provide quantum symmetries for the Hamiltonian   $\hcH_N$ (\ref{b3}) by imposing the condition
  \be
 \bigl [\hI_N,\hcH_N\bigr]= \bigl [\hI_N',\hcH_N\bigr]=0 ,
  \label{b6}
  \ee
which determines the explicit form for the monomials $\hP_\ell^{(s)}$ together with the values of $A_\ell^{(s)}$ and $B_\ell^{(s)}$. Provided that these quantum symmetries have been found, it follows that the sets $(\hcH_N,\hcC,\hI_N )$ and $(\hcH_N,\hcC,\hI'_N )$ are formed each by three algebraically independent operators, so that $\hcH_N$ is a quantum superintegrable Hamiltonian. As happens in the classical case (see~\cite{zk2023}), the four operators $(\hcH_N,\hcC,\hI_N ,\hI'_N)$ satisfy an algebraic dependence relation. 

The second step is to compute the corresponding spectrum, which must exhibit   a maximal degeneracy of the energy levels, by means of an algebraic procedure. This requires the obtainment of the associated deformed oscillator algebra~\cite{Daskaloyannis1991,Quesne1994,Quesne1997,Plyushchay1997,Quesne2000,Kumar,Carballo,Rebesh,Lee}; for the 
use of deformed oscillator algebras in superintegrability and their connection with the symmetry algebra we refer to~\cite{Daskaloyannis2001,Quesne2007,Marquette2009,Marquette2013,MarquetteQuesne2013} and references therein.
 In our case, for a given $N$, this step can be systematized as follows:
\begin{itemize}

\item  Define the following operators from the quantum symmetries  $(\hcC,\hI_N ,\hI'_N)$:
\be
 \hcK_1:=\hcC , \qquad 
 \hcK_2:= \frac{1}{2}(\hI_N'-\hI_N) ,\qquad 
 \hcK_3:= [ \hcK_1, \hcK_2] ,
\label{b7}
\ee
which close on a polynomial deformation of $\mathfrak{sl}(2,\mathbb R)\simeq \mathfrak{so}(2,1)$.

\item Introduce  number and ladder (raising/lowering) operators $( \hcK,\hcK_+,\hcK_-)$ from $(\hcK_1,\hcK_2,\hcK_3)$ via a nonlinear change of basis
(which depends on the chosen $N$ through the parameters $\gamma_k$), such that
\be
\bigl[\hcK,\hcK_\pm\bigr]=\pm\hcK_\pm,\qquad \bigl[\hcK_-,\hcK_+\bigr]=\Phi\bigl(\hcH_N,\hcK+  {\rm Id} \bigr)-\Phi\bigl(\hcH_N,\hcK  \bigr)  ,
\label{b8}
\ee
where ${\rm Id}$ is the identity operator and $\Phi$ is a structure function. The latter is assumed to factorize as 
\be
\hcK_+\hcK_-\equiv \Phi\bigl(\hcH_N,\hcK\bigr)=\Phi_1\bigl(\hcH_N,\hcK\bigr)\Phi_2\bigl(\hcH_N,\hcK\bigr).
\label{b9}
\ee

\item The deformed oscillator algebra arises in a basis $(\hB,\hbm,\hbp)$ by setting
\be
 \hB:= \hcK- u\,  {\rm Id},\qquad
 \hbm:=\hcK_- ,\qquad
\hbp:=\hcK_+,
\label{b10}
\ee
 where  $u$ is a constant. The commutation relations read as
 \be
 \bigl[\hB,\hbp\bigr]=\hbp,\quad\ \bigl[\hB,\hbm\bigr]=-\hbm,\quad\  \bigl[\hbm,\hbp\bigr]=\Phi\bigl(\hcH_N,\hB+(u+1) {\rm Id}\bigr)-\Phi\bigl(\hcH_N,\hB+ u\,{\rm Id}\bigr),
 \label{b11}
 \ee
where $\hbp\hbm=\Phi\bigl(\hcH_N,\hB+ u\,{\rm Id}\bigr)$.

\item Finally, using the eigenvalues $E$ of $\hcH_N$ and $\B$ of $\hB$, we consider a  finite-dimensional   representation of (\ref{b11}), that is,
\be
\Phi=\Phi(\B,E,u)=\Phi_1(\B,E,u)\Phi_2(\B,E,u),
   \label{b12b}
\ee
subjected to the conditions 
\begin{equation}
  \Phi(0,E,u)=0,\qquad \Phi(\n+1,E,u)=0 
   \label{b12}
\end{equation}  
for a natural number $\n\in\{1,2,\dots\}$.
This set of algebraic equations  provides us with constraints on the spectrum and the representation-dependent constant $u$. Hence the solutions take the form $u=u(\n)$ and $E=E(\n)$, thus leading to $ \Phi(\B,E(\n),u(\n))\equiv \Phi(\B,\n)$ for $\B\in\{ 1,\dots,\n\}$. If  $\Phi(B,\n)>0$, the finite-dimensional representation is also unitary.  The  parameter $\n$, which characterizes the dimension of the subspace, {\em i.e.}~of dimension $ \n + 1$, determines the principal quantum number $\fn$.

\end{itemize}

 In the following sections~\ref{s3} and \ref{s4}, we work out the quadratic and cubic generalized quantum Zernike Hamiltonians $\hcH_N$ (\ref{b3}), corresponding to the values $N=2$ and $N=3$, respectively.
For each case, we obtain the quantum symmetries (\ref{b6}), the polynomial algebras (\ref{b7})  and (\ref{b8}), together with the deformed oscillator algebra (\ref{b11}). From the latter, the solutions for the possible spectra determined by the equations~(\ref{b12}) are deduced and analyzed. Furthermore, such new results for $N=3$ are  also interpreted as superintegrable perturbations of the isotropic oscillator on the  two-dimensional sphere, hyperbolic and Euclidean spaces; recall that the usual isotropic Euclidean oscillator is just the case with $N=1$, thus with a single parameter $\gamma_1$.

In addition, following a similar construction, the fourth-order with $N=4$ and the fifth-order with $N=5$ systems are explicitly solved. To facilitate the reading of the paper, the corresponding results are presented in Appendices~\ref{appendixA} and  \ref{appendixB}, respectively. 
While we have not yet fully derived the general expressions for the deformed oscillator algebra (\ref{b11})  along with the corresponding spectra for arbitrary $N$, 
we present two conjectures in section \ref{s5} that address these problems.  These conjectures serve as a foundation for future work to establish their proofs.

%%%%%%%%%%%%%%%%%%%%%%%%%%%%%%%%%%%%%%%%%%%%%%

 \sect{The proper (quadratic) quantum Zernike Hamiltonian}
 \label{s3}

Let us consider the quadratic case of  $\hcH_N$ (\ref{b3}) with $N=2$ and arbitrary parameters $\gamma_1$ and $\gamma_2$.
The second-order in the momenta quantum symmetries (\ref{b6b}) commuting with  $\hcH_2$ (\ref{b6}) are found to be
\be
\begin{split}
  \hI_2 &= \hp_2^2+ \gamma_1\hq_2\hp_2 -\gamma_2\bigl( \hq_2^2( \hp_1^2-\hp_2^2) -2 \hq_1\hq_2\hp_1\hp_2 +\ri  \hq_1\hp_1+\ri  \hq_2\hp_2\bigr)  \\
  &= \hp_2^2+ \gamma_1\hq_2\hp_2      +\gamma_2\bigl(   (\hq_1^2+\hq_2^2  ) \hp_2^2  -\hcC^2 \bigr)    ,\\[2pt]
    \hI'_2 &= \hp_1^2+\gamma_1\hq_1\hp_1+\gamma_2\bigl(\hq_1^2+\hq_2^2 \bigr) \hp_1^2,  \\
\end{split}
\label{c1}
\ee
where $\hcC$ is the quantum angular momentum (\ref{b5}). These satisfy the relation
\be
\hcH_2 =   \hI_2+  \hI_2'.
\label{c2}
\ee
From $(\hcC,\hI_2 ,\hI'_2)$ we introduce the operators $(\hcK_1,\hcK_2,\hcK_3)$ defined in (\ref{b7}), which obey the following polynomial commutation relations:
\be
\begin{split}
\bigl[ \hcK_1,\hcK_2 \bigr]&= \hcK_3, \qquad  \bigl[ \hcK_1,\hcK_3 \bigr]= 4   \hcK_2 -2 \gamma_2    \hcK_1^2   ,\\[2pt]
\bigl[ \hcK_2,\hcK_3 \bigr]&=    \bigl(  \gamma_1^2 +2\ri  \gamma_1\gamma_2  +2 \gamma_2 \hcH_2 \bigr) \hcK_1+ 4\gamma_2 \hcK_1 \hcK_2-  2\gamma_2 \hcK_3  .
\end{split}
\label{c3}
\ee
We observe that the corresponding classical functions $I_2$, $I_2'$ and $\cK_i$'s coming from the quantum operators
(\ref{c1}) and (\ref{b7})  do not  exactly coincide  with $\mathcal I_2$, $\mathcal I_2'$ and $\mathcal L_i$'s as obtained in~\cite{zk2023}. Actually, they are related to these through  $I_2=\mathcal I_2-\gamma_2 \cC^2 $, $I_2'=\mathcal I_2'$,   where $\cC$, $\mathcal I_2\equiv \mathcal I$ and $\mathcal I'_2\equiv \mathcal I'$  are given in (\ref{a5}),  (\ref{a5b}) and  (\ref{a5c}), respectively and, therefore,  they are constants of the motion of the classical Hamiltonian $\cH_2 $   (\ref{a4}).  This explains why the classical counterpart of the quantum algebra (\ref{c3}) does not provide us with the Poisson symmetry  algebra determined by the functions $\mathcal L_i$'s in~\cite{zk2023},  given by formulae (\ref{a5e}) and  (\ref{za5fd}).
  These slight differences are also observable for higher values of $N$, and have their origin in the presence or absence of the quantum angular momentum operator $\hcC$ in the quantum symmetries.

We define the number and ladder operators $( \hcK,\hcK_+,\hcK_-)$ in (\ref{b8}) as
\be
\begin{split}
\hcK&:= \frac{1}{2 } \hcK_1,\\
\hcK_+&:=   \hcK_2+\frac 12  \hcK_3- \frac{1}{2}\gamma_2  \hcK_1^2,\\
\hcK_-&:=  \hcK_2-\frac 12 \hcK_3- \frac{1}{2}\gamma_2  \hcK_1^2,
\end{split}
\label{c5}
\ee
which close on the polynomial symmetry algebra given by
\be
\bigl[ \hcK,\hcK_\pm \bigr]=\pm \hcK_\pm ,\qquad \bigl[ \hcK_-,\hcK_+ \bigr]=2  \bigl(  \gamma_1^2 + 2 \ri \gamma_1\gamma_2 +2\gamma_2\hcH_2 \bigr) \hcK+16\gamma_2^2 \hcK^3 .
\nonumber
\ee
The symmetry algebra of the quantum Hamiltonian  $\hcH_2$  becomes, over this basis, a cubic polynomial algebra $\mathfrak{sl}^{(3)}(2,\mathbb R)$, which is just the so-called Higgs algebra~\cite{Higgs} (see also~\cite{Debergh2000,Debergh2003,Ruan2003,Civitarese2007,Kepler2009} and references therein), as already pointed out in~\cite{PSWY2017}. 

In order to construct a deformed Heisenberg algebra and compute the corresponding spectrum of  $\hcH_2$, we need to find the structure function $\Phi$ in (\ref{b8}), which turns out to be
 \be
\begin{split}
\hcK_+\hcK_-=\Phi(\hcH_2,\hcK)&=\frac{1}{4}\bigl( 4 \gamma_2 -2\ri \gamma_1  +\hcH_2\bigr)\hcH_2
-   \bigl( \gamma_1^2 + 2 \ri \gamma_1\gamma_2   + 2\gamma_2  \hcH_2\bigr)\hcK\\
&\qquad  + \bigl(\gamma_1^2+4\gamma_2^2    + 2 \ri \gamma_1\gamma_2   + 2\gamma_2  \hcH_2\bigr)\hcK^2-8\gamma_2^2 \hcK^3+4\gamma_2^2  \hcK^4.
\end{split}
\nonumber
\ee
Following (\ref{b9}), the factorization of $\Phi$ is given by
\be
\begin{split}
\Phi(\hcH_2,\hcK)&=\Phi_1(\hcH_2,\hcK)\Phi_2(\hcH_2,\hcK),\qquad \bigl[\Phi_1(\hcH_2,\hcK),\Phi_2(\hcH_2,\hcK)\bigr]=0,\\[2pt]
\Phi_1(\hcH_2,\hcK) &=\frac 14 \bigl( \hcH_2 - 2\ri \gamma_1 \hcK +4 \gamma_2 \hcK^2\bigr),\\[2pt]
\Phi_2 (\hcH_2,\hcK)&=  \hcH_2 - 2 (\ri \gamma_1-2 \gamma_2 ) {\rm Id} + 2 (\ri \gamma_1-4 \gamma_2 ) \hcK +4 \gamma_2 \hcK^2 \\
&= \hcH_2 + 2\ri \gamma_1\bigl(\hcK  - {\rm Id} \bigr) + 4\gamma_2\bigl(\hcK  - {\rm Id} \bigr)^2,
\end{split}
\label{c8}
\ee
which  gives  rise to the deformed oscillator algebra (\ref{b11})  under the change of operators (\ref{b10}).

%%%%%%%%%%%%%%%%%%%%%%%%%%%%%%%%%%%%%%%%%%%%%%%%%%%

 \subsect{Spectrum  of the quadratic system}
 \label{s31}
 
At this stage, we have all the required information to compute the spectrum $E$ of  $\hcH_2$ in (\ref{b4}) algebraically. Using (\ref{c8}), it follows that the finite-dimensional representation of (\ref{b11}) yields the structure function $\Phi$ (\ref{b12b}) as
\be
\begin{split}
\Phi(\B,E,u) &=\Phi_1(\B,E,u)\Phi_2(\B,E,u),\\
\Phi_1(\B,E,u) &=\frac 14  \bigl( E - 2\ri \gamma_1 (\B+u) +4 \gamma_2  (\B+u)^2\bigr),\\
\Phi_2(\B,E,u) &= E + 2  \ri \gamma_1 (\B+u-1)  +4 \gamma_2   (\B+u-1)^2 ,
\end{split}
\label{c9}
\ee
 where we recall that  $E$  and $\B$ are the eigenvalues of $\hcH_2$ and   $\hB$, respectively. Taking into account the two constraints in (\ref{b12}), the result can be summarized in the following 
  
\begin{proposition} 
\label{prop1}
The set of equations (\ref{b12}) yields  four types of solutions for the representation-dependent constant $u$ and the spectrum of the Hamiltonian $\hcH_2$ in (\ref{b4})  depending on the  parameter  $\n\in\{1,2,\dots\}$, that is,   $u=u(\n)$ and   $E=E(\n)$. Introducing these expressions into (\ref{c9}),   the corresponding   structure functions $\Phi(\B,E(\n),u(\n))\equiv \Phi(\B,\n)$ for $\B=\{1,\dots,\n\}$ are obtained. The final result is displayed in Table~\ref{table1}.
\end{proposition}

 %%%%%%%%%%%%%%%%%%%%%%%%%%%%%%%%

  %%%%%%%%%%%%%%%% TABLE 1%%%%%%%%%%%%%%%%%%

\begin{table}[t]
 
{\small
\caption{ \small{The four types of solutions of the equations (\ref{b12}) with the corresponding representation dependent constant $u(\n)$, spectrum $E(\n)$ and structure function $\Phi(B,\n)$ (\ref{c9}) where $\n\in\{1,2,\dots\}$ and $B\in\{1,\dots,\n\}$.}}
  \begin{center}
\noindent 
\begin{tabular}{  l l l}
\hline

\hline
\\[-6pt]
Type {\rm I} & \multicolumn{2}{l}{ $\displaystyle   u_{\rm I} = -\frac \n2$ \qquad $   \displaystyle  E_{\rm I}= -   \n (\ri \gamma_1+\gamma_2   \n)$} \\[6pt]
 & \multicolumn{2}{l}{$\Phi_{\rm I}=-  \B(\B-\n-1) \bigl(\ri \gamma_1 + 2 \gamma_2   (\B-1) \bigr)
  \bigl(\ri \gamma_1 - 2 \gamma_2   (\B-\n) \bigr)$}\\ [6pt]

 Type {\rm II} &  \multicolumn{2}{l}{ $\displaystyle   u_{\rm II} = -\frac \n 2$\qquad   $   \displaystyle  E_{\rm II}=    (\n +2)\bigr(\ri \gamma_1-\gamma_2   (\n+2) \bigl)$} \\[6pt]
 & \multicolumn{2}{l}{$\Phi_{\rm II}=- \B(\B-\n-1) \bigl(\ri \gamma_1 - 2 \gamma_2   (\B+1) \bigr)
  \bigl(\ri \gamma_1 + 2 \gamma_2   (\B-\n-2) \bigr)$}\\[6pt] 

Type {\rm III} & $\displaystyle   u_{\rm III} = -\frac{1}{2}\left( \n-1+\frac{\ri \gamma_1}{2\gamma_2} \right)$ & $\   \displaystyle  E_{\rm III}=    - \frac{  \gamma_1^2}{4\gamma_2} - \gamma_2 (\n +1)^2$ \\[8pt]
 & \multicolumn{2}{l}{$\Phi_{\rm III}= \B(\B-\n-1) \bigl(\ri \gamma_1 - 2 \gamma_2   (\B+1) \bigr)
  \bigl(\ri \gamma_1 - 2 \gamma_2   (\B-\n) \bigr)$}\\ [6pt] 
   
Type {\rm IV} & $\displaystyle   u_{\rm IV} = -\frac{1}{2}\left( \n+1-\frac{\ri \gamma_1}{2\gamma_2} \right)$ & $\   \displaystyle  E_{\rm IV}=    - \frac{  \gamma_1^2}{4\gamma_2} - \gamma_2 (\n +1)^2$ \\[8pt]
 & \multicolumn{2}{l}{$\Phi_{\rm IV}= \B(\B-\n-1) \bigl(\ri \gamma_1 + 2 \gamma_2   (\B-1) \bigr)
  \bigl(\ri \gamma_1 + 2 \gamma_2   (\B-\n-2) \bigr)$}
 \\[10pt]
 \hline

\hline
\end{tabular}
 \end{center}
\label{table1}
}
\end{table}

%%%%%%%%%%%%%%%%%%%%%%%%%%%%%%%%%%%%%%%%%%%%%%%%%%%

  Some remarks  are in order.   It can be directly verified that the four types of solutions in table~\ref{table1} satisfy both equations (\ref{b12}), so that the factors $\Phi_1$ and $\Phi_2$  in (\ref{c9}) cancel out as follows:
   \be
\begin{array}{ll}
\Phi_1(0,E_{\rm I},u_{\rm I})= \Phi_2(n+1,E_{\rm I},u_{\rm I})=0,\quad & \Phi_1(n+1,E_{\rm II},u_{\rm II})= \Phi_2(0,E_{\rm II},u_{\rm II})=0, \\[4pt]
\Phi_2(0,E_{\rm III},u_{\rm III})= \Phi_2(n+1,E_{\rm III},u_{\rm III})=0,\quad & \Phi_1(0,E_{\rm IV},u_{\rm IV})= \Phi_1(n+1,E_{\rm IV},u_{\rm IV})=0 .
 \end{array}
\nonumber   
\ee  
  The solutions of type III and IV give rise to the same spectrum, $E_{\rm III}=E_{\rm IV}$, although the structure functions $\Phi$ and hence the deformed oscillator algebras in (\ref{b11}) are different. The four types of spectra take real values whenever the parameter $\gamma_1$ is a pure imaginary number, keeping $\gamma_2$ as a real parameter. This suggests to introduce arbitrary real parameters $\alpha$ and $\beta$  as defined in (\ref{a6}), and appearing within the Zernike Hamiltonian (\ref{b2}). With this convention, we find that
  \be
E_{\rm I}= -   \n (  \beta+\alpha   \n) ,\qquad E_{\rm II}=    (\n +2)\bigr(\beta -\alpha (\n+2) \bigl),\qquad E_{\rm III}=  E_{\rm IV}=    \frac{  \beta^2}{4\alpha} -\alpha (\n +1)^2.
   \label{c11}
  \ee
  Different possibilities appear depending on   the specific values of $\alpha$ and $\beta$. 
 In this regard, it should be noted that if the value $\alpha= \gamma_2=0$ is allowed, the spectrum $E_{\rm III}=  E_{\rm IV}$ is not well-defined. This fact is quite important, since this case corresponds to the system with $N=1$, which actually corresponds to the isotropic oscillator on the Euclidean plane, that we study next. By contrast, the two spectra $E_{\rm I}$ and $E_{\rm II}$ do cover the system with $\alpha= \gamma_2=0$. Moreover, we also point out that  $E_{\rm II}$ can be related to  $E_{\rm I}$ by applying a `shift' $n\to (n+2)$ along with the change $\beta\to -\beta$ to $E_{\rm I}$.

Furthermore,  and quite unexpectedly, the quantum Zernike system corresponding to the values $\alpha=-1$ and $\beta=-2$ arises as a very `special' case, as the four types of solutions merge into the same, namely
  \be
E (\n)= \n(\n+2),\qquad u(\n)=  -\frac \n 2,\qquad \Phi(\B,\n) = 4\B^2(\B-\n-1)^2, 
 \nonumber
  \ee
  where $\n\in\{1,2,\dots\}$ and $\B\in\{1,\dots,\n\}$. As $\Phi(\B,n)>0$ holds, there is no restriction on the value of $\n$, and the finite-dimensional representation is always unitary. Identifying  the parameter  $\n$ with the principal quantum number $\fn$, we recover the expression 
    \be
  E=\fn(\fn+2) 
     \label{c13}
     \ee
given in~\cite{PSWY2017}   ({\em cf.}~eq.~(16)). Observe that, strictly speaking, $\n\in\{1,2,\dots\}$, while $\fn\in\{0,1,2,\dots\}$, indicating that the difference relies merely on the ground state.

  %%%%%%%%%%%%%%%%%%%%%%%%%%%%%%%%%%%%%%%%%%%%%%%%%%%

 \subsect{Application to the spherical and hyperbolic oscillators}
 \label{s32}

 In addition, it is worth remarking that, by means of a canonical transformation, it has been shown in~\cite{zk2023} (see also~\cite{Fordy2018}) that  the classical Zernike system $\cH_{\rm Zk} \equiv \cH_{2} $ (\ref{b3}) can be regarded as an isotropic oscillator on the two-dimensional  sphere $\mathbf S^2$  and   the hyperbolic space $\mathbf H^2$.  
 Within this interpretation, the 
imaginary coefficient  $\gamma_1$ is related to the frequency $\omega$ in the form $\gamma_1=  2\ri \omega$, while the second one is just minus the constant curvature of the space $\gamma_2=-\kappa$. 
For the sake of completeness, we recall the explicit form of  $ \cH_{2} $ as a natural Hamiltonian in terms of a curved  kinetic energy $\mathcal T_\kappa$ and a central potential $\mathcal U_\kappa\propto \tan^2(\sqrt{\kappa}\rho) /\kappa$:
 \be
\cH_{2} =\mathcal T_\kappa+\mathcal U_\kappa(\rho) ,\qquad \mathcal T_\kappa =  p_\rho^2 + \frac{\kappa
\, p_\phi^2 } {  \sin^2(\sqrt{\kappa}  \rho)  }  , \qquad \mathcal U_\kappa(\rho)= -\frac{ \gamma_1^2 }{4\kappa}  \tan^2(\sqrt{\kappa}\rho)  ,
\label{hamgeodesic}
 \ee
 where $\rho$ is a geodesic radial coordinate, {\em i.e.}, the distance between the particle and the origin on the space measured along the geodesic joining both points, while $\phi\in[0,2\pi)$ is an ordinary angle.
  Therefore, $\cH_{2} $ covers the well-known two-dimensional isotropic spherical or Higgs oscillator  on $\mathbf S^2$  ($\gamma_2<0$, $\kappa>0$),  the hyperbolic system on  $\mathbf H^2$ ($\gamma_2>0$,  $\kappa<0$) and the usual one on the Euclidean plane $\mathbf E^2$  $(\gamma_2=-\kappa=0)$ (see~\cite{Santander,2013Higgs,2014Higgs,Kuruannals, vero} and references therein). For the `standard' values of the curvature $\kappa\in\{+1,0,-1\}$ we find that (\ref{hamgeodesic}) gives
  \be
\begin{array}{lll}
\mathbf S^2& (\kappa=+1)\!\!:&  \quad \displaystyle 
\cH_{2} =   p_\rho^2 + \frac{ 
\, p_\phi^2 } {  \sin^2\! \rho }  -\frac{ \gamma_1^2 }{4 }  \tan^2\!\rho   .\\[10pt]
\mathbf E^2& (\kappa=0)\!\!:&  \quad \displaystyle 
\cH_{2} =   p_\rho^2 + \frac{ 
\, p_\phi^2 } {    \rho^2 }  -\frac{ \gamma_1^2 }{4 }\,  \rho^2   .\\[10pt]
\mathbf H^2& (\kappa=-1)\!\!:&  \quad \displaystyle 
\cH_{2} =   p_\rho^2 + \frac{ 
\, p_\phi^2 } {  \sinh^2\! \rho }  -\frac{ \gamma_1^2 }{4 }  \tanh^2\!\rho   .
 \end{array}
\nonumber   
\ee
   From this perspective, the solutions of type III and IV in Table~\ref{table1} are not well-defined for the flat Euclidean case with $\kappa=-\gamma_2\to 0$, as we have already mentioned.

Although both quantum curved spherical and hyperbolic oscillators have already been fully  solved in~\cite{Santander,Kuruannals}, obtaining the corresponding spectrum and eigenfunctions, our algebraic procedure allows us to recover their spectrum in straightforward manner, that shows deep differences between the spherical and hyperbolic cases. Nevertheless, both reproduce the Euclidean oscillator under the flat limit $\kappa\to 0$,   which is only allowed for types I and II in Table~\ref{table1}.

Let us focus on the solution of type I in Table~\ref{table1}, setting $\gamma_1=2\ri$ {\em i.e.}~$\beta=-2$   (this means to fix   $\omega=1$) and express $\gamma_2=\alpha=-\kappa$, allowing any value for the curvature $\kappa$. We analyze separately the three particular systems according to $\kappa$ and, for simplicity, hereafter we drop the index `I'.
 
 The case with $\kappa=0$ leads to the isotropic oscillator on  $\mathbf E^2$   (hence, the case with $N=1$ in (\ref{b3})), such that
 \be
 E (n)= 2n,\qquad  \Phi =4  \B(\n+1-\B),
 \label{ee2}
 \ee
with $\n\in\{1,2,\dots\}$, $\B=\{1,\dots,\n\}$ so that $\Phi>0$. Thus, as expected,   the spectrum is linear in $\n$, unbounded and with a constant difference  $ E (n+1)- E (n)=2$. This implies that there exists an infinite set of bound states.

 The spherical or Higgs oscillator arises on $\mathbf S^2$ when $\kappa>0$, leading to the expressions
 \be
E (n)= 2n+\kappa n^2,\qquad  \Phi =4  \B(\n+1-\B) \bigr(1+\kappa (B-1) \bigl)  \bigr(1+\kappa (n-B) \bigl),
\label{es2}
 \ee
 where, again, $\n\in\{1,2,\dots\}$, $\B=\{1,\dots,\n\}$ and $\Phi>0$. Now the spectrum is quadratic in $\n$, with higher values than the Euclidean oscillator, and unbounded.
  Nevertheless, the difference of the energy between two consecutive states is no longer constant but increases linearly with $\n$ as
 \be
 E (n+1)-E(n)=2 +\kappa(2n+1).
 \label{difs}
 \ee
 We conclude that, in this case, there  also exists an infinite number of bound states. We represent the first energies (\ref{es2}) for three values of the curvature, and those of the Euclidean oscillator (\ref{ee2}) in figure~\ref{fig1}. As the curvature grows, the differences between the spherical and Euclidean spectra become larger.

For the hyperbolic oscillator on $\mathbf H^2$ we write $\kappa=-|\kappa|$, obtaining that
\be
E (n)= 2n-|\kappa| n^2,\qquad  \Phi =4  \B(\n+1-\B) \bigr(1-|\kappa| (B-1) \bigl)  \bigr(1-|\kappa|(n-B) \bigl).
\label{eh2}
 \ee
In contrast to (\ref{difs}), the difference of the energies is decreasing and becomes
 \be
 E (n+1)-E(n)=2 -|\kappa|(2n+1).
  \label{difh}
 \ee
 In order to relate these results to those previously deduced in~\cite{Santander,Kuruannals}  and recover  the Euclidean system by a limiting process,  we require  a positive energy spectrum $E(n)$     (\ref{eh2}) and positive differences of two consecutive energies (\ref{difh}). The first condition leads to $2>|\kappa| n$, thus the lower value $n=1$ determines  a limiting   value  for the curvature: $|\kappa|<2$.  The second 
 requirement   implies that  there exists a maximum value $n_{\rm max}$ for the positive integer $n\ge 1$,   which is  the greatest integer number ensuring  that the relation  (\ref{difh}) is always positive, namely
 $$
  E (n_{\rm max})-E(n_{\rm max}-1)=2 -|\kappa|\big(2(n_{\rm max}-1)+1\big)>0.
$$
 This constraint gives rise to
    \be
   1\le   n_{\rm max}<\frac 1{|\kappa|}+\frac 12   .
     \label{constraints}
  \ee
Notice that the  single state (ground case)  for the extreme value $  n_{\rm max}=1$ is also included, since it consistently  implies that $|\kappa|<2$.  Moreover, observe that the specific values appearing in relations (\ref{eh2})--(\ref{constraints})  depend not only  on the (negative) curvature, but also on  the  particular choice of the parameter $\beta=-2$   (and, therefore, on the frequency $\omega$). And if $|\kappa|\to 0$, then $n_{\rm max}\to\infty$, recovering the Euclidean case.

Let us now prove that the restriction (\ref{constraints}) always ensures that   $\Phi>0$ in (\ref{eh2}), showing that   $ \bigr(1-|\kappa| (B-1) \bigl)>0$ and $  \bigr(1-|\kappa|(n-B) \bigl)>0$. Taking into account that   $\B=\{1,\dots,\n\}$, the maximum value for $B\ge 1$ is just $ n_{\rm max}$ (\ref{constraints}).  For the first factor we find that
$$
\frac 1{|\kappa|}+1> n_{\rm max}\ge B \ge 1.
$$
And for the second factor we take the extreme values $n=n_{\rm max}$ and $B=1$:
$$
\frac 1{|\kappa|}>n-B\quad \Longleftrightarrow\quad \frac 1{|\kappa|}>n_{\rm max}-1.
$$

 % \newpage
%%%%%%%%%%% figure1 %%%%%%%%%%%%%%%%%%

 \begin{figure}[t]
\centering
\includegraphics[scale=0.55]{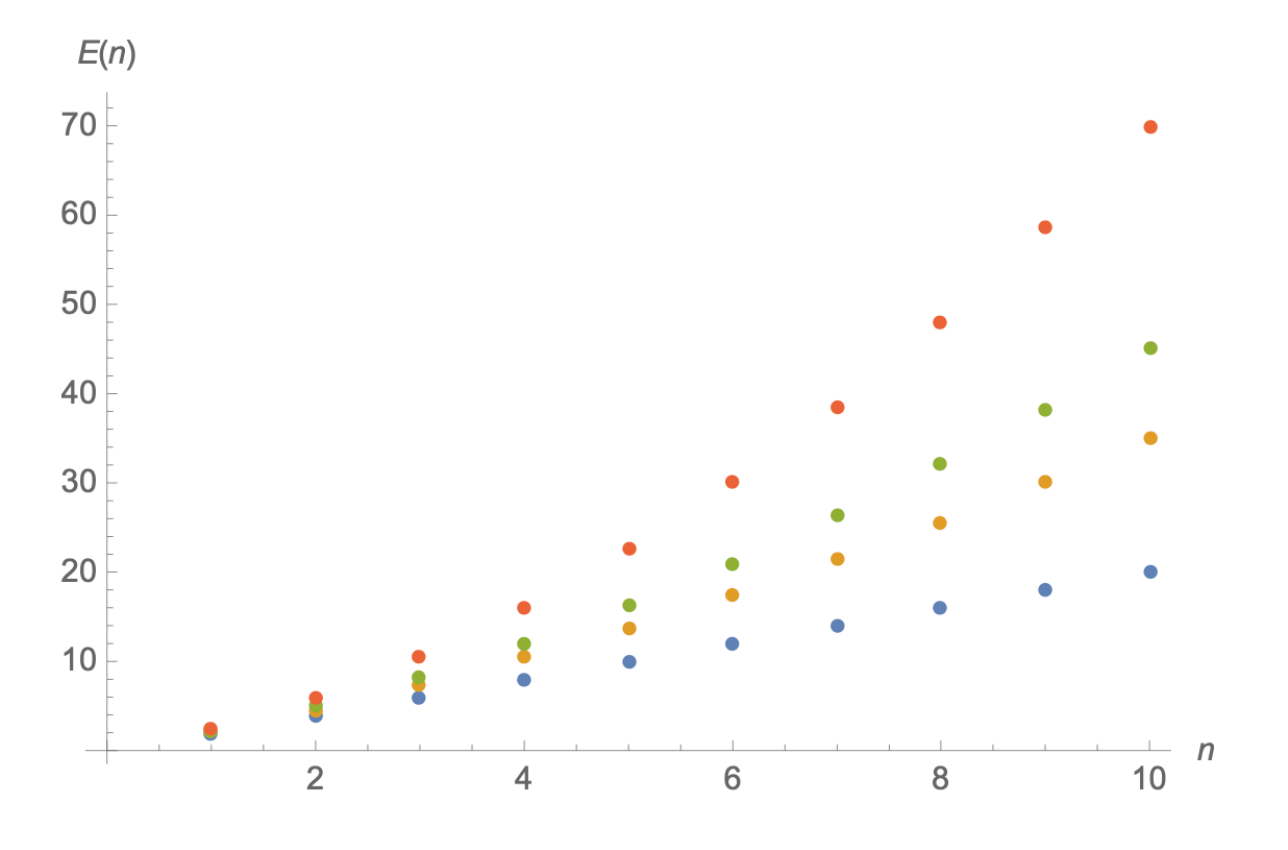}
\caption{\small The  discrete spectrum (\ref{es2}) of the spherical oscillator on $\mathbf S^2$ for the first ten states $1\le n\le 10$  according to  three values of  the curvature $\kappa\in\{0.5,\, 0.25,\, 0.15\}$ starting from the upper dots together with the Euclidean spectrum  (\ref{ee2}) with $\kappa=0$ in the lowest values.}
\centering
     \label{fig1}
\end{figure}

%%%%%%%%%%%%%%%%%%%%%%%%%%%%%%%%%%%%%%%%%%%%%%

    %%%%%%%%%%% figure2 %%%%%%%%%%%%%%%%%%

 \begin{figure}[htp!
]
\centering
\includegraphics[scale=0.455]{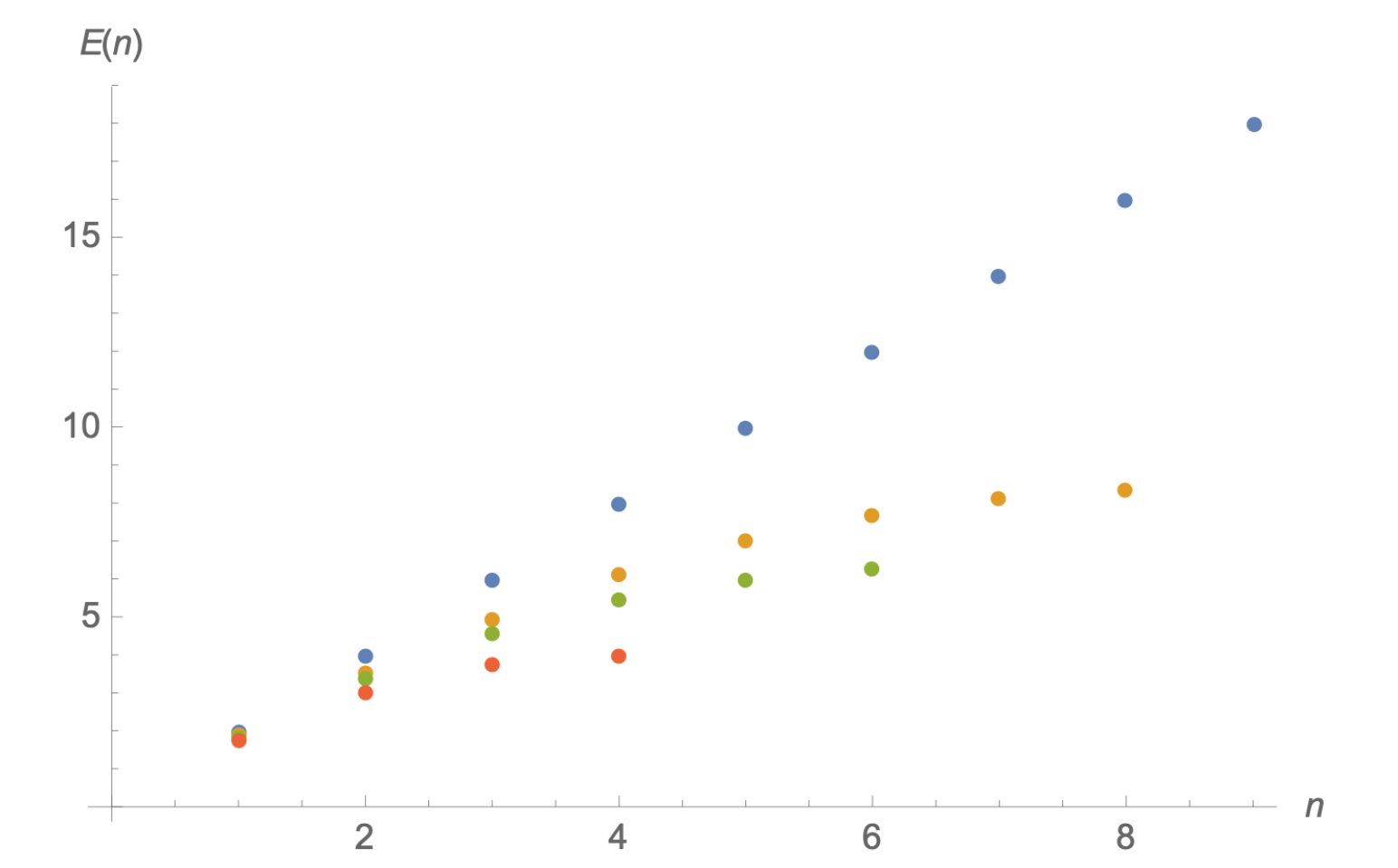}
\caption{\small The finite discrete spectrum (\ref{eh2}) of the hyperbolic oscillator on $\mathbf H^2$ for the    three values of  the curvature $\kappa\in\{-0.25,\, -0.16,\, -0.12\}$ corresponding to 
 $n_{\rm max}\in\{4,\, 6,\, 8\}$ (\ref{constraints})
 starting from the lower dots, respectively, together with the Euclidean spectrum  (\ref{ee2}) with $\kappa=0$ in the highest values.}
\centering
     \label{fig2}
\end{figure}

%%%%%%%%%%%%%%%%%%%%%%%%%%%%%%%%%%%%%%%%%%%%%%

   Therefore, in agreement with~\cite{Santander,Kuruannals},    the  isotropic hyperbolic oscillator  has  a positive  finite discrete spectrum    with $\n\in\{1,\dots,n_{\rm max}\}$, and there exist  unbounded states which may be provided by one-dimensional hyperbolic P\"oschl--Teller potentials, as pointed out in Theorem 6 in~\cite{Kuruannals}.    
 We illustrate these results  in figure~\ref{fig2}, plotting the discrete spectra  for three values of $n_{\rm max}$ (so choosing a value for the curvature) which are   always lower than
   the  Euclidean ones (\ref{ee2}).

%\newpage
%%%%%%%%%%%%%%%%%%%%%%%%%%%%%%%%%%%%%%%%%%%%%%

 \sect{The cubic quantum Zernike Hamiltonian}
 \label{s4}

In this section we study the quantum symmetries and deduce the possible spectra for the cubic Hamiltonian  $\hcH_3$ (\ref{b3})  depending on three arbitrary parameters $\gamma_1$,  $\gamma_2$ and $\gamma_3$. It is worthy to be observed that the cubic  $\gamma_3$-term in $\hcH_3$ can be interpreted as a superintegrable perturbation of the quantum Zernike Hamiltonian $\hcH_{\rm Zk}$ in (\ref{b2}). Indeed, we require that all the expressions for  $\hcH_2$ in (\ref{b3}) obtained in the previous section are recovered whenever $\gamma_3$ is put equal to zero.

The two quantum symmetries (\ref{b6b}) of  $\hcH_3$, that are of third-order in the momenta, are obtained after some computation, and adopt the form 
\be
\begin{split}
  \hI_3 &=  \hp_2^2+ \gamma_1\hq_2\hp_2      +\gamma_2\bigl(   (\hq_1^2+\hq_2^2  ) \hp_2^2  -\hcC^2 \bigr)  \\[2pt]
   &\qquad +\gamma_3\bigl(  \hq_2^3( \hp_2^3- \hp_1^2\hp_2)+( \hq_1^3+ 3 \hq_1\hq_2^2 ) \hp_1  \hp_2^2- 3 \ri \hq_2^2\hp_2^2- 3 \ri \hq_1 \hq_2\hp_1\hp_2-\hq_2  \hp_2\bigr)  ,    \\[4pt]
    \hI'_3 &= \hp_1^2+\gamma_1\hq_1\hp_1+\gamma_2\bigl(\hq_1^2+\hq_2^2 \bigr) \hp_1^2\\[2pt]
    &\qquad +\gamma_3\bigl(  \hq_1^3( \hp_1^3- \hp_1\hp_2^2)+( \hq_2^3+ 3 \hq_1^2 \hq_2)  \hp_1^2\hp_2 - 3 \ri \hq_1^2\hp_1^2- 3 \ri \hq_1 \hq_2\hp_1\hp_2-\hq_1  \hp_1\bigr).   
\end{split}
\nonumber
\ee
They satisfy the relation (\ref{c2}). The operators $(\hcK_1,\hcK_2,\hcK_3)$, as defined in (\ref{b7}), close on the polynomial algebra given by
\be
\begin{split}
\bigl[ \hcK_1,\hcK_2 \bigr]&= \hcK_3, \qquad  \bigl[ \hcK_1,\hcK_3 \bigr]= 4   \hcK_2 -2 \gamma_2    \hcK_1^2   ,\\[2pt]
\bigl[ \hcK_2,\hcK_3 \bigr]&=    \bigl(  \gamma_1^2 +2\ri  \gamma_1\gamma_2  -4  \gamma_1\gamma_3  +2 (\gamma_2+3\ri \gamma_3 )\hcH_3 \bigr) \hcK_1+ 4\gamma_2 \hcK_1 \hcK_2-  2\gamma_2 \hcK_3 \\
&\qquad - 4\gamma_3  ( \gamma_1-\ri \gamma_2-\gamma_3     )\hcK_1^3+ 3\gamma_3^2\hcK_1^5.
\end{split}
\nonumber
\ee
The number and ladder operators $( \hcK,\hcK_+,\hcK_-)$ are  defined exactly as in (\ref{c5}),  and satisfy the commutation relations (\ref{b8}) with 
 \be
 \begin{split}
  \bigl[ \hcK_-,\hcK_+ \bigr]&=2  \bigl(  \gamma_1^2 + 2 \ri \gamma_1\gamma_2 -4  \gamma_1\gamma_3  +2 (\gamma_2+3\ri \gamma_3 )\hcH_3   \bigr) \hcK\\ 
  &\qquad +16\bigl( \gamma_2^2+2 \gamma_3^2-2  \gamma_1 \gamma_3+2 \ri \gamma_2 \gamma_3 \bigr)\hcK^3 + 96\gamma_3^2\hcK^5.
 \end{split}
\nonumber
\ee
The symmetry algebra of the cubic quantum Hamiltonian  $\hcH_3$  hence corresponds to a fifth-order polynomial Higgs-type algebra  $\mathfrak{sl}^{(5)}(2,\mathbb R)$ (see~\cite{zk2023} for the classical picture).

On the other hand, the structure function $\Phi$ (\ref{b8}) is determined by  
 \be
\begin{split}
\hcK_+\hcK_-&=\Phi(\hcH_3,\hcK)\\
&=\frac{1}{4}\bigl( 4 \gamma_2 -2\ri \gamma_1  + 8\ri \gamma_3+\hcH_3\bigr)\hcH_3
-   \bigl( \gamma_1^2 + 2 \ri \gamma_1\gamma_2 -4 \gamma_1\gamma_3  + 2(\gamma_2 +3 \ri \gamma_3 )\hcH_3\bigr)\hcK\\
&\qquad  + \bigl(\gamma_1^2+4\gamma_2^2    + 2 \ri \gamma_1\gamma_2   -12\gamma_1\gamma_3+8\ri \gamma_2\gamma_3    + 2(\gamma_2 +3 \ri \gamma_3 )\hcH_3\bigr)\hcK^2\\[2pt]
&\qquad -8\bigl( \gamma_2^2+2 \gamma_3^2-2  \gamma_1 \gamma_3+2 \ri \gamma_2 \gamma_3 \bigr)\hcK^3+4\bigl(  \gamma_2^2+12  \gamma_3^2 -2  \gamma_1 \gamma_3+2 \ri \gamma_2 \gamma_3\bigr) \hcK^4\\[2pt]
&\qquad -48 \gamma_3^2\hcK^5+16  \gamma_3^2\hcK^6 ,
\end{split}
\nonumber
\ee
which is then factorized  as (see (\ref{b9}))
\be
\begin{split}
\Phi(\hcH_3,\hcK)&=\Phi_1(\hcH_3,\hcK)\Phi_2(\hcH_3,\hcK),\qquad \bigl[\Phi_1(\hcH_3,\hcK),\Phi_2(\hcH_3,\hcK)\bigr]=0,\\
\Phi_1(\hcH_3,\hcK) &=\frac 14 \bigl( \hcH_3 - 2\ri \gamma_1 \hcK +4 \gamma_2 \hcK^2+ 8\ri \gamma_3 \hcK^3\bigr),\\[2pt]
\Phi_2 (\hcH_3,\hcK)&=  \hcH_3 - 2 (\ri \gamma_1-2 \gamma_2 -4\ri \gamma_3) {\rm Id} + 2 (\ri \gamma_1-4 \gamma_2 -12 \ri \gamma_3) \hcK \\
&\qquad +4( \gamma_2+6\ri \gamma_3) \hcK^2-8\ri \gamma_3 \hcK^3 \\
&=   \hcH_3+ 2\ri \gamma_1\bigl(\hcK  - {\rm Id} \bigr) + 4\gamma_2\bigl(\hcK  - {\rm Id} \bigr)^2-
 8\ri \gamma_3\bigl(\hcK  - {\rm Id} \bigr)^3.
\end{split}
\nonumber
\ee
It is routine to verify that with the change of operators (\ref{b10}), we get the deformed oscillator algebra (\ref{b11}).

%%%%%%%%%%%%%%%%%%%%%%%%%%%%%%%%%%%%%%%%%%%%%%%%%%%

 \subsect{Spectrum  of the cubic system}
 \label{s41}

In terms of the finite-dimensional   representation of  the deformed oscillator algebra (\ref{b11}), the structure function $\Phi$ of (\ref{b12b}) adopts the  expression
\be
\begin{split}
\Phi(\B,E,u) &=\Phi_1(\B,E,u)\Phi_2(\B,E,u),\\
\Phi_1(\B,E,u) &=\frac 14  \bigl( E - 2\ri \gamma_1 (\B+u) +4 \gamma_2  (\B+u)^2 +8\ri  \gamma_3  (\B+u)^3\bigr),\\
\Phi_2(\B,E,u) &= E + 2  \ri \gamma_1 (\B+u-1)  +4 \gamma_2   (\B+u-1)^2 -8\ri  \gamma_3  (\B+u-1)^3.
\end{split}
\label{d6}
\ee
 The solution can thus be summarized in the following statement.
  
%%%%%%%%%%%%%%%%%%%%%%%%%%%%%%%%
 
  \begin{proposition} 
\label{prop2}
(i) The set of equations (\ref{b12}) coming from (\ref{d6}) yields  the corresponding solutions for the representation-dependent constant $u=u(\n)$ and the spectrum $E=E(\n)$ of the Hamiltonian $\hcH_3$ in (\ref{b4})  depending on the parameter  $\n\in\{1,2,\dots\}$. They lead to the structure function  $\Phi(\B,E(\n),u(\n))\equiv \Phi(\B,\n)$ for $\B=\{1,\dots,\n\}$ via the factorization (\ref{d6}).\\
(ii) The requirement that the solutions are well-defined for any values of $\gamma_1$ and $\gamma_2$ for $\gamma_3=0$ 
reduces the set of solutions to only two, given by 
\be
\begin{array}{lll}
\!\!\!\!\mbox{\rm Type I:}& \displaystyle   u_{\rm I} = -\frac \n 2,&\quad \displaystyle  E_{\rm I}= -    \bigl(\ri \gamma_1 \n+\gamma_2   \n^2 - \ri \gamma_3 \n^3\bigr), \\[8pt]
&  \multicolumn{2}{l}{  \Phi_{\rm I}=-  \B(\B-\n-1) \bigl(\ri \gamma_1 + 2 \gamma_2   (\B-1) +\ri \gamma_3\bigr(\n(2\B-\n-2)-4(\B-1)^2 \bigl)\bigr)  }\\[6pt]
  &  \multicolumn{2}{l}{\qquad\qquad \times \bigl(\ri \gamma_1 - 2 \gamma_2   (\B-\n) +\ri \gamma_3\bigr(3 \n(2\B-\n)-4\B^2 \bigl) \bigr).}\\[4pt]
\!\!\!\! \mbox{\rm Type II:}&  \displaystyle   u_{\rm II} = -\frac \n 2,&\quad E_{\rm II}=     \ri \gamma_1 (\n +2)-\gamma_2   (\n+2)^2  - \ri \gamma_3 (\n+2)^3 ,  \\[8pt]
&  \multicolumn{2}{l}{ 
\Phi_{\rm II}=- \B(\B-\n-1) \bigl(\ri \gamma_1 - 2 \gamma_2   (\B+1)  +\ri \gamma_3\bigr(\n(2\B-\n-2)-4( \B^2+\B+1) \bigl)\bigr) }\\[6pt]
   &  \multicolumn{2}{l}{\qquad\qquad \times   \bigl(\ri \gamma_1 + 2 \gamma_2   (\B-\n-2) +\ri \gamma_3\bigr(3(\n+2)(2\B-\n-2)-4\B^2 \bigl)\bigr) .}
\end{array}
\nonumber   
\ee
\end{proposition}

 %%%%%%%%%%%%%%%%%%%%%%%%%%%%%%%%

To illustrate the situation with the  discarded solutions for the limit $\gamma_3\to 0$, let us consider, for instance, one of them given by 
\be
u=-\frac \n 2-\frac{1}{2\gamma_3}\sqrt{\gamma_3\bigl( \gamma_1-2\ri \gamma_2 \n - 3 \gamma_3 \n^2\bigr)} \, ,\qquad E=-\frac 1{\gamma_3}\bigl(\gamma_2-2\ri \gamma_3 \n\bigr)\bigl( \gamma_1-2\ri \gamma_2 \n - 4\gamma_3 \n^2 \bigr).
\label{d8}   
\ee
For $\gamma_1\neq 0$ or $\gamma_2\neq 0$, it is easily seen that $u$, $E$  or both are not well-defined for $\gamma_3=0$.

The two consistent types of solutions given in proposition~\ref{prop2} constitute a cubic extension of those presented in proposition~\ref{prop1} and displayed in Table~\ref{table1}. We observe that types III and IV do not admit such a cubic generalization, implying that they are specific solutions for the quadratic Hamiltonian (\ref{b3})  with $N=2$  (although they are not well-defined if $\gamma_2=0$). The spectrum corresponding
to   types I and II takes real values whenever the coefficients $\gamma_1$ and $ \gamma_3$
are pure imaginary numbers, while $\gamma_2$ is real. By considering (\ref{a6}), as before, we denote
\be
\gamma_1=-\rm i \beta,\qquad \gamma_2=\alpha ,\qquad \gamma_3=\ri \mmu, \qquad \{\alpha,\beta,\mmu\}\in\mathbb R,
\label{d9}
\ee
finding that
\be
\begin{split}
E_{\rm I}&= -    \bigl(\beta \n+\alpha   \n^2 +\mmu \n^3\bigr),\\
E_{\rm II}&=  \beta (\n +2)-\alpha   (\n+2)^2  +\mmu (\n+2)^3 ,
\end{split}
\label{d10}
\ee
to be compared with (\ref{c11}). Furthermore, if we set $\alpha=-1$ and $\beta=-2$ and identify $\n$ 
 with the principal quantum number $\fn$, we obtain two possible superintegrable perturbations of the spectrum (\ref{c13}) of  the Zernike Hamiltonian $\hcH_{\rm Zk}$  in (\ref{a1}), reading as
    \be
E_{\rm I}=\fn(\fn+2) -\mmu \fn^3,\qquad E_{\rm II}=\fn(\fn+2) +\mmu(\fn+2)^3.
\nonumber
     \ee

 %%%%%%%%%%%%%%%%%%%%%%%%%%%%%%%%%%%%%%%%%%%%%%%%%%%

 \subsect{Perturbations of the  Euclidean,  spherical and hyperbolic oscillators}
 \label{s42}

 Following the analysis in section~\ref{s32}, we can interpret the results of proposition~\ref{prop2} as superintegrable cubic perturbations of the isotropic oscillator on $\mathbf E^2$, $\mathbf S^2$ and $\mathbf H^2$. To this end, let us consider the solution of type I (\ref{d10}) (eliminating  the index `I'), again setting $\beta=-2$ and $ \alpha=-\kappa$, but now keeping an arbitrary real parameter $\mu$, which characterizes the perturbation.  Thus, we have six possibilities, according to the curvature $\kappa$ and the two signs of $\mu$.
 
 For the  flat Euclidean oscillator  with $\kappa=0$,   we start with a negative value for the perturbation parameter $\mu=-|\mu|$, obtaining that
 \be
E (n)= 2n+|\mu| n^3,\qquad E (n+1)-E(n)=2 +|\mu|(3n^2+3n+1).
 \label{s4a}
 \ee
 Hence, the spectrum is cubic in $n$, unbounded and  the energy between two consecutive states grows quadratically with $\n$. It is worth noting that the behaviour of the spectrum (\ref{s4a}) is quite similar to that of the spherical or Higgs oscillator (\ref{es2}) and (\ref{difs}), so we can say that it is a {\em cubic spherical perturbation} of the Euclidean oscillator. In figure~\ref{fig3} we plot the first energies (\ref{s4a}) for three negative values of $\mu$ together with those of the non-perturbed Euclidean oscillator (\ref{ee2}) (to be compared with figure~\ref{fig1}). 
   
 Then, we take a positive value for $\mu$  on $\mathbf E^2$ finding that
 \be
E (n)= 2n- \mu n^3,\qquad E (n+1)-E(n)=2 - \mu(3n^2+3n+1).
 \label{s4b}
 \ee
We require a positive spectrum and positive differences of two consecutive energies.  
The condition $E (n) >0$ implies that $\mu n^2<2$, giving rise to  a limit value for the perturbation parameter: $\mu  <2$. The second  requirement yields a maximum value $n_{\rm max}$ for the positive integer $n\ge 1$, similarly to the hyperbolic oscillator (see (\ref{constraints})):
\begin{align}
E (n_{\rm max})-E(n_{\rm max}-1)&=2 - \mu\big(3(n_{\rm max}-1)^2+3(n_{\rm max}-1)+1\big) \nonumber\\
& = 2 - \mu\big(3 n_{\rm max}^2-3n_{\rm max}+1\big) >0 .
\nonumber
\end{align}
From this, we find that $n_{\rm max}$
is  the greatest integer number that satisfies the following constraint
 \be
    1\le   n_{\rm max}<\frac 12+ \frac 16\sqrt{ \frac {24}{\mu} - 3} \,.
     \label{s4c}
 \ee 
 Note that if we consider the ground state $n_{\rm max}=1$, we recover that  $\mu  <2$.
In this respect, we can say that (\ref{s4b}) and (\ref{s4c}) determine a   {\em cubic hyperbolic perturbation} of the Euclidean oscillator that leads to a positive finite discrete  spectrum with $\n\in\{1,\dots,n_{\rm max}\}$. We represent  the discrete spectrum  for three values of $n_{\rm max}$ (so for $\mu$)  in  figure~\ref{fig4},  which are  lower than  the  Euclidean ones (\ref{ee2}), and  worthy of comparison with figure~\ref{fig2}.

%\newpage

%%%%%%%%%%% figure3 %%%%%%%%%%%%%%%%%%

 \begin{figure}[t]
\centering
\includegraphics[scale=0.465]{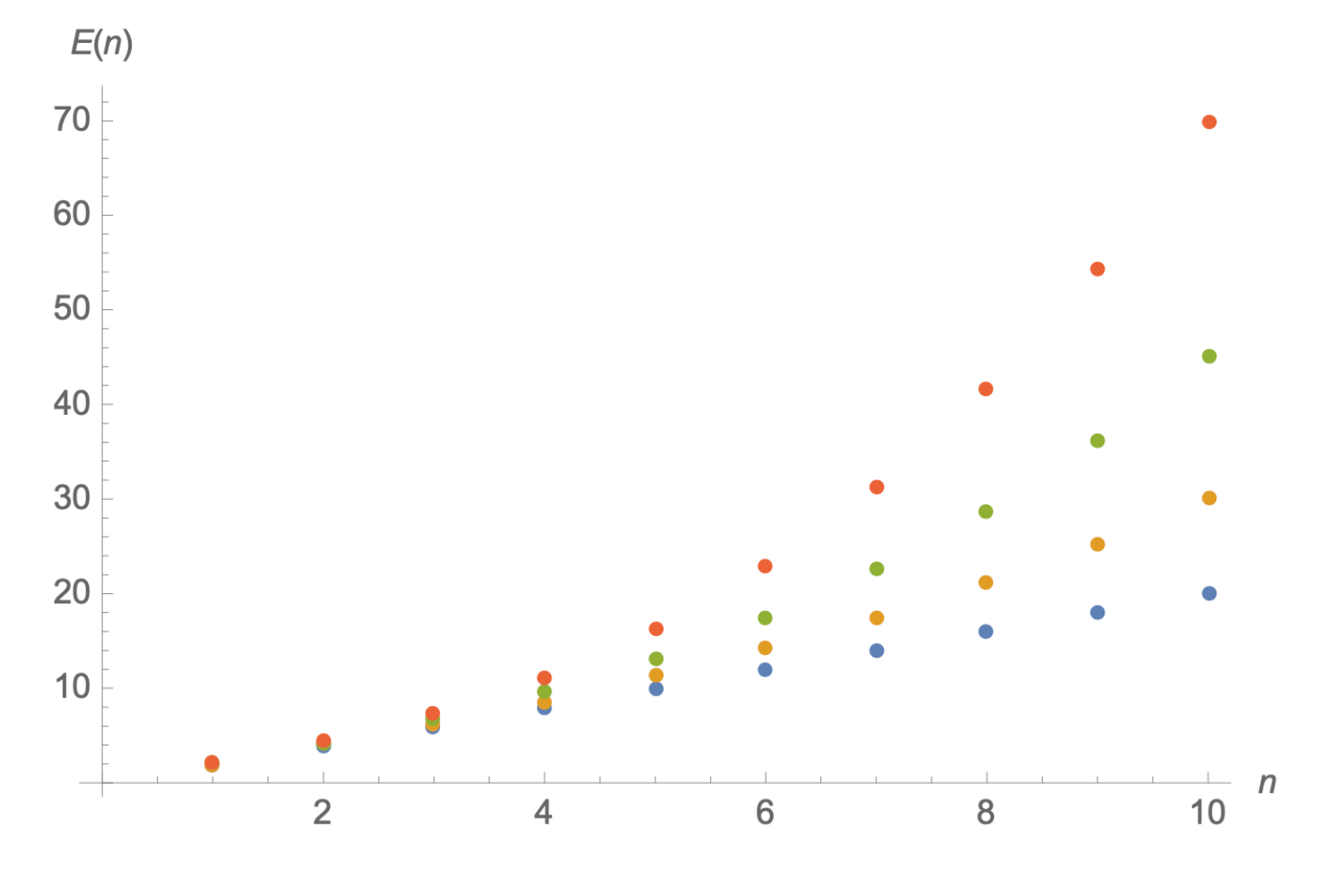}
\caption{\small The  discrete spectrum (\ref{s4a}) of the cubic spherical perturbation of the oscillator on $\mathbf E^2$ for the first ten states $1\le n\le 10$  according to  three values of  the parameter $\mu\in\{-0.05,\, -0.025,\, -0.01\}$ starting from the upper dots together with the Euclidean spectrum  (\ref{ee2}) with $\mu=0$ in the lowest values.}
\centering
     \label{fig3}
\end{figure}

%%%%%%%%%%%%%%%%%%%%%%%%%%%%%%%%%%%%%%%%%%%%%%

    %%%%%%%%%%% figure4 %%%%%%%%%%%%%%%%%%

 \begin{figure}[htp!
]
\centering
\includegraphics[scale=0.40]{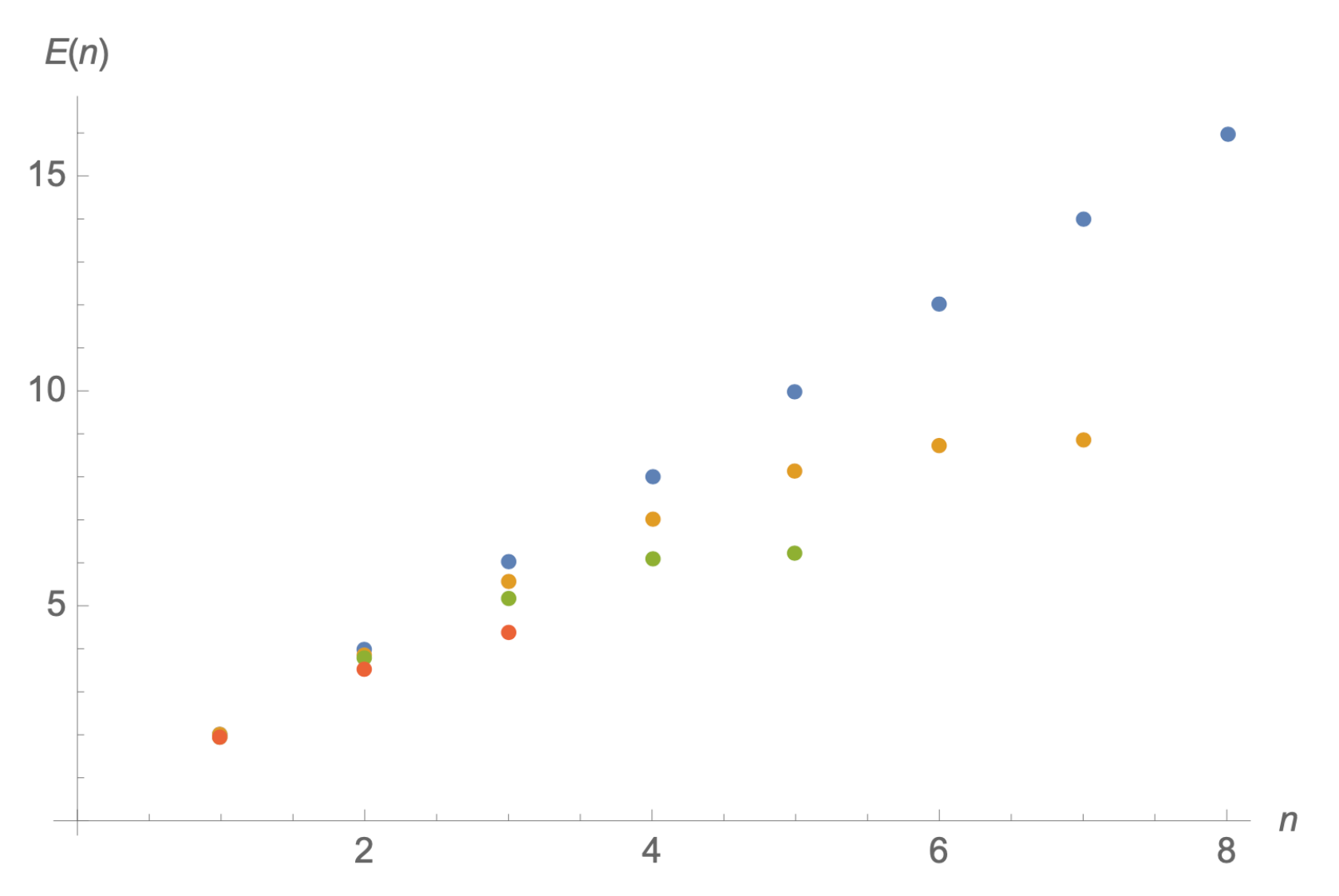}
\caption{\small The finite discrete spectrum (\ref{s4b}) of the cubic hyperbolic perturbation of the oscillator on $\mathbf E^2$ for      three values  $\mu\in\{0.06,\, 0.03,\, 0.015\}$ corresponding to 
 $n_{\rm max}\in\{3,\, 5,\, 7\}$ (\ref{s4c})
 starting from the lower dots, respectively, together with the Euclidean spectrum  (\ref{ee2}) with $\mu=0$ in the highest values.}
\centering
     \label{fig4}
\end{figure}

%%%%%%%%%%%%%%%%%%%%%%%%%%%%%%%%%%%%%%%%%%%%%%
%\newpage

Now we consider the sphere $\mathbf S^2$ with positive curvature $\kappa>0$ and a negative parameter  $\mu=-|\mu|$. We arrive at the expression
  \be
E (n)= 2n+\kappa  n^2+|\mu| n^3,\qquad E (n+1)-E(n)=2 +\kappa(2n+1)+|\mu|(3n^2+3n+1).
\nonumber
 \ee
 As a consequence, the spectrum is discrete and unbounded, but the presence of $\mu<0$ entails a higher energy growth than in the spherical oscillator (\ref{es2}) and (\ref{difs}), such that it can be regarded as  a {\em cubic spherical perturbation}  of the spherical-Higgs oscillator.
 
 The above situation changes drastically when $\mu>0$. Actually, in this case
   \be
E (n)= 2n+\kappa   n^2-\mu n^3,\qquad E (n+1)-E(n)=2 +\kappa(2n+1)-\mu(3n^2+3n+1).
 \label{s4f}
 \ee
  The requirement to obtain a positive spectrum leads to the restriction $\mu<2+\kappa$, 
while   imposing   positive differences of two consecutive energies implies that there is  a maximum value $n_{\rm max}$ for the positive integer $n\ge 1$ according to the   inequality 
$$
E (n_{\rm max})-E(n_{\rm max}-1)  = 2+\kappa (2n_{\rm max}-1 ) - \mu\big(3 n_{\rm max}^2-3n_{\rm max}+1\big) >0 ,
$$
thus arriving at
 \be
    1\le   n_{\rm max}<\frac 12+\frac{\kappa}{3\mu}+ \frac 16\sqrt{ \frac {4\kappa^2+24\mu}{\mu^2} - 3}\ .
     \label{s4g}
 \ee
 Notice that if $n_{\rm max}=1$, then we recover that $\mu<2 +\kappa$.

In this way, the expressions (\ref{s4f}) and (\ref{s4g}) characterize   a  {\em cubic hyperbolic perturbation} of the spherical oscillator that leads to a positive finite discrete  spectrum with $\n\in\{1,\dots,n_{\rm max}\}$. Observe that the unbounded discrete spectrum   (\ref{es2}) is recovered under the  limit $\mu\to 0$ (so 
$n_{\rm max}\to \infty$). The flat limit $\kappa\to 0$ is also well-defined, reproducing the 
hyperbolic perturbation of the Euclidean oscillator determined by (\ref{s4b}) and  (\ref{s4c}).
Figure~\ref{fig5} shows  the discrete spectra  for several values of $n_{\rm max} $ (so for $\mu$), which can be compared with figure~\ref{fig1} ($\mu=0$) and figure~\ref{fig4}  ($\kappa=0$).

Likewise, both types of perturbations can be applied to the oscillator on the hyperbolic space  $\mathbf H^2$ with negative curvature $\kappa= - |\kappa|<0$ (see (\ref{eh2})--(\ref{constraints}));   these are  the spherical perturbation for $\mu=-|\mu|<0$ and the hyperbolic perturbation for $\mu>0$. A similar analysis to the one previously performed for the sphere $\mathbf S^2$ would be necessary for each type of perturbation. However, the results are quite cumbersome, as there are several possibilities for $n_{\rm max}$ depending on $|\kappa|$ and $\mu$. For brevity in the exposition, these explicit computations are omitted.

 %%%%%%%%%%% figure5 %%%%%%%%%%%%%%%%%%
 
 \begin{figure}[t]
\centering
\includegraphics[scale=0.55]{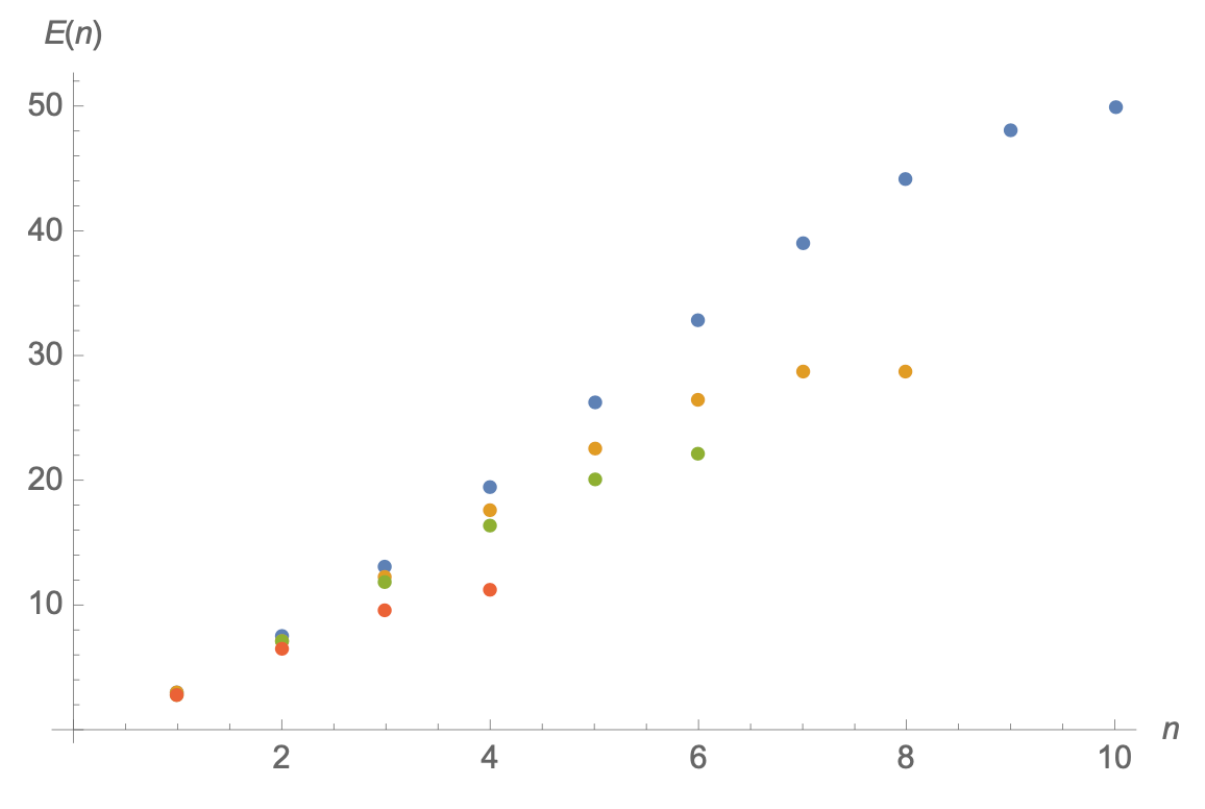}
\caption{\small The finite discrete spectrum (\ref{s4f}) of the cubic hyperbolic perturbation of the spherical oscillator on $\mathbf S^2$  with $\kappa=+1$ for      $\mu\in\{0.2,\, 0.12,\, 0.1,\, 0.07\}$ corresponding to 
 $n_{\rm max}\in\{4,\, 6,\, 8,\, 10\}$ (\ref{s4g})
 starting from the lower dots.}
\centering
     \label{fig5}
\end{figure}

%%%%%%%%%%%%%%%%%%%%%%%%%%%%%%%%%%%%%%%%%%%%%%

\newpage
 
 %%%%%%%%%%%%%%%%%%%%%%%%%%%%%%%%%%%%%%%%%%%%%%%%%%%
 \sect{Higher-order quantum Zernike Hamiltonians: Conjectures}
 \label{s5}
  
 So far,   the quantum Hamiltonian $\hcH_N$ (\ref{b3}) has been analyzed in detail for the values $ 1\le N\le 3$. Following the same procedure, the cases with $N=4$ and $N=5$  can be explicitly solved, which are presented in Appendices~\ref{appendixA} and~\ref{appendixB}, respectively.
However,  a closed explicit expression for the higher-order quantum symmetries $\hI_N$ and $\hI'_N$ for arbitrary values of $N$ has not been found, due to the exponentially increasing computational difficulties. Nevertheless, based on the results for the values analyzed, $1\le N\le 5$, the form of the structure function $\Phi(\hcH_N,\hcK)$ in (\ref{b8}) along with its factorizing terms (\ref{b9}) can be inferred, allowing us to deduce the two types I and II of solutions for the spectrum that generalize those 
already obtained.  This extrapolation is presented in form of conjectures, as follows.

 \begin{conjecture} 
  \label{conj1} (i) For any $N\ge 1$ and any value of the  coefficients $\gamma_k$ ($1\leq k\leq N)$, the quantum Hamiltonian $\hcH_N$  (\ref{b3}) admits three quantum symmetries corresponding to the quantum angular momentum operator $\hcC$ in (\ref{b5}) and two operators $\hI_N$ and $\hI'_N$ (\ref{b6b}), which  are quadratic in the momenta for $N=1,2$ and of $N^{{\rm th}}$-order for $N>2$.\\
  (ii) The sets $\big (\hcH_N,\hcC, \hI_N\big)$ and $\big (\hcH_N,\hcC, \hI'_N\big )$ are  formed by three algebraically independent operators, hence $\hcH_N$ determines a superintegrable system.\\
  (iii) The set $\big ( \hcC, \hI_N,\hI'_N\big )$ leads to the quantum symmetries $\big(\hcK_1,\hcK_2,\hcK_3\big)$ in (\ref{b7}) and, in consequence,  it is possible to define the number and ladder operators $\big( \hcK,\hcK_+,\hcK_-\big)$ verifying the commutation relations (\ref{b8}) with the structure function $\Phi= \Phi_1 \Phi_2$ (\ref{b9}), whose factorizing terms are given by 
\be
\begin{split}
&\Phi_1\bigl(\hcH_N,\hcK\bigr)=\frac 14 \left(\hcH_N -\sum_{k=1}^N  (2\ri)^k \gamma_k\hcK^k\right) ,\\
&\Phi_2\bigl(\hcH_N,\hcK\bigr)= \hcH_N -\sum_{k=1}^N  (-2\ri)^k\gamma_k \bigl( \hcK -{\rm Id} \bigr)^k .
\end{split}
\label{f1}
\ee
\end{conjecture}

Observe that   $[\Phi_1 ,\Phi_2 ]=0$ and that  the operators   $\big( \hcK,\hcK_+,\hcK_-\big)$ should span a $(2N-1)^{{\rm th}}$-order polynomial symmetry algebra of   Higgs-type   $\mathfrak{sl}^{(2N-1)}(2,\mathbb R)$ determined by the commutator $[\hcK_-,\hcK_+]$.

Provided that the conjecture~\ref{conj1} is satisfied, we can further deduce the factorizing terms in \eqref{b9} (hence the structure function $\Phi$) and the solutions consistent with the generalization of the quantum Zernike Hamiltonian.

 \begin{conjecture} 
  \label{conj2} (i) The structure function $\Phi(\B,E,u)=\Phi_1\Phi_2$ (\ref{b12b})  is obtained  from  (\ref{f1})  by introducing the  finite-dimensional   representation of   (\ref{b11}), namely 
   \be
\begin{split}
&\Phi_1(\B,E,u)=\frac 14 \left(E -\sum_{k=1}^N  (2\ri)^k \gamma_k(B+u)^k\right) ,\\
&\Phi_2(\B,E,u)= E-\sum_{k=1}^N  (-2\ri)^k \gamma_k ( B+u - 1  )^k .
\end{split}
\label{f2}
\ee
(ii)   The set of equations (\ref{b12}) coming from (\ref{f2}) gives rise to two solutions   for the representation dependent constant $u=u(\n)$ and the spectrum $E=E(\n)$ of the Hamiltonian $\hcH_N$~(\ref{b4})  depending on the parameter  $\n\in\{1,2,\dots\}$, that hold for any values of the coefficients $\gamma_k$. From these solutions the structure function  $\Phi(\B,E(\n),u(\n))\equiv \Phi(\B,\n)=\Phi_1(\B,\n)\Phi_2(\B,\n)$ for $\B=\{1,\dots,\n\}$ is obtained using (\ref{f2}). The resulting expressions read as
\be
\begin{array}{lll}
 \mbox{\rm Type I:}& \displaystyle   u_{\rm I} = -\frac \n 2,&\quad \displaystyle  E_{\rm I}= \sum_{k=1}^N  (-\ri)^k \gamma_kn^k , \\[6pt]
 &  \multicolumn{2}{l}{ \displaystyle \Phi_{\rm I}=  \frac 14 \left( \sum_{k=1}^N  (-\ri)^k \gamma_k
 \left(n^k-(n-2B)^k \right)  \right) \left( \sum_{k=1}^N  (-\ri)^k \gamma_k
 \left(n^k-(2B-n-2)^k \right)  \right) .}\\[12pt]
 \mbox{\rm Type II:}&  \displaystyle   u_{\rm II} = -\frac \n 2,&\quad\displaystyle E_{\rm II}=   \sum_{k=1}^N  \ri^k \gamma_k(n+2)^k,\\[6pt]
 &  \multicolumn{2}{l}{ \displaystyle \Phi_{\rm II}=   \frac 14  \left( \sum_{k=1}^N  \ri^k \gamma_k
 \left((n+2)^k-(2B-n)^k \right)  \right) \left( \sum_{k=1}^N  \ri^k \gamma_k
 \left((n+2)^k-(n+2-2B)^k \right)  \right) .}
 \end{array}
\nonumber
\ee
 \end{conjecture} 
 
 Some observations concerning conjecture~\ref{conj2} seem pertinent. First of all, 
 it can be easily checked that both types of solutions satisfy the equations (\ref{b12}), as
 $$
 \Phi_1(0,E_{\rm I},u_{\rm I})= \Phi_2(n+1,E_{\rm I},u_{\rm I})=0,\qquad  \Phi_1(n+1,E_{\rm II},u_{\rm II})= \Phi_2(0,E_{\rm II},u_{\rm II})=0.
 $$
 Second, to ensure that both spectra $E_{\rm I}$ and $E_{\rm II}$ are real, it is necessary to require that $\gamma_k$ is purely imaginary for odd values of $k$, while it must be real for even values of $k$.  Third,  the above results are well-defined for any value of all the coefficients $\gamma_k$. In particular, we can take a single or more coefficients different from zero. For instance, the simplest choice is to set only $\gamma_1\ne 0$,   {\em i.e.}, $N=1$, leading to a linear spectrum in $n$, that corresponds to the first system with a discrete spectrum; in fact, it has been interpreted as the isotropic Euclidean oscillator in section~\ref{s32}
 (when $\gamma_k=0$ for all $k$, we recover the trivial geodesic motion on the Euclidean plane). Finally, from this point of view, the generalized quantum Zernike Hamiltonian $\hcH_N$ in (\ref{b3})   can be regarded as the superposition of $N$ possible potentials, each of them determined by one parameter $\gamma_k\ne 0$, leading to a term of $k^{{\rm th}}$-order in the momenta in the quantum symmetries  $\hI_N$ and $\hI'_N$, as well as a term $n^k$  (or $(n+2)^k$) in the spectrum.

In summary, in order to consistently prove the conjectures \ref{conj1}  and \ref{conj2}, it is first necessary to find the generic expression of the quantum symmetries  $\hI_N$ and $\hI'_N$ in (\ref{b6b}), from which the structure of the polynomial Higgs-type algebra of order $(2N-1)^{{\rm th}}$ is obtained,   namely $\mathfrak{sl}^{(2N-1)}(2,\mathbb R)$, which constitutes the main computational obstruction. A second step consists in finding the precise structure of the quantum operators $( \hcK,\hcK_+,\hcK_-)$, starting from $(\hcK_1,\hcK_2,\hcK_3)$ in (\ref{b7}), leading to the factorization terms in (\ref{f1}). As already shown for $1\le N\le 5$, this is far from being a trivial task, even with the use of symbolic computer packages.

    %%%%%%%%%%%%%%%%%%%%%%%%%%%%%%%%%%%%%%%%%%%%%%%%%%%

\sect{Conclusions and outlook}
\label{s6}

In this work, a quantum version of the generalized classical Zernike Hamiltonian, first studied in~\cite{zk2023}, has been proposed, showing that one of its main features, namely the superintegrability property, is preserved by quantization. In addition, it has been shown that the spectrum of the quantum Hamiltonian \eqref{b3} can be inferred by purely algebraic means, using the quantum symmetries and the structure of the associated polynomial Higgs-type algebra. Although explicit computations become extremely cumbersome for values $N\geq 6$, a generic prescription has been given to solve the spectrum problem for any given value of $N$. Based on the general shape of the factorizing terms of the structure function and the resulting spectra for the admissible solutions that allow us to interpret the result as a higher-order perturbation of the usual quantum Zernike model, we conjecture the solutions for arbitrary value  of $N$, without making explicit use of the quantum symmetries, for which the exact expression is still an open problem. In this context, another unanswered question concerns the number of solutions of equation \eqref{b12}, and whether for higher values of $N$ we can find additional solutions that are well-defined and behave properly  for the limits $\gamma_k\rightarrow 0$ for $k\geq 3$, albeit this seems not probable in view of the structure observed for values $N\leq 5$.    

Furthermore, we stress that the interpretation of the Hamiltonian $\cH_{2} $ (\ref{b3}) as defining, in a unified form, the isotropic oscillator on $ \mathbf E^2$, $\mathbf S^2$ and $\mathbf H^2$, have allowed us to consider the generalized Zernike systems as superintegrable perturbations for these oscillators when $N\ge 3$. Moreover, we have also shown that such perturbations can be of spherical or hyperbolic type, depending on the value of the corresponding coefficient $\gamma_k$. In this sense, these new results extend Bertrand's theorem to the momentum-dependent systems studied here.

In any case, the main open problem is to solve the corresponding Schr\"odinger equation (\ref{b4})  by obtaining both eigenfunctions and spectra, and then analyze them with respect to the algebraic results. This task is quite arduous, as demonstrated by the solutions for $N=2$, already carried out in~\cite{Santander,Kuruannals}.

 There is a number of interesting problems that emerge from our analysis, and that deserve a more profound investigation. In particular:
\begin{itemize}

\item The Schr\"odinger equation could be faced for the simplest flat Euclidean system,  corresponding to  $N=1$, at least for   the cubic  $N=3$ perturbation, and  the resulting wave eigenfunctions and spectrum analyzed according to the spherical or hyperbolic type of the perturbation.  This result can provide the path to solve the Schr\"odinger equation for the perturbations of curved oscillators on $\mathbf S^2$ and $\mathbf H^2$.

\item  The generalized quantum Zernike Hamiltonians could be implemented in the three Lorentzian spaces of constant curvature, Minkowski, anti-de Sitter and de Sitter spacetimes, by means of graded contractions or analytical continuation~\cite{SIGMA2006}; the appropriate coordinates would be the geodesic polar ones shown in (\ref{hamgeodesic}).

\item As higher-order momentum-dependent potentials have already been considered  in quantum molecular and nuclear dynamics (see~\cite{DDR1987,BG1988,CH1995momentumpotentials,LGXZL2003,NMS2020} and references therein), it would be natural to search for some application of the generalized Zernike systems in these frameworks.

\end{itemize}

Other possible open problems arise when considering the underlying 
$\mathfrak{gl}(2)$-coalgebra symmetry of the Hamiltonian $\cH_{N} $ (\ref{b3}), for any $N$, which we briefly summarize.

 Let us consider the Lie algebra $\mathfrak{gl}(2)=\spn\{J_-,J_+,J_3,\Xi \}$  with commutation relations and Casimir operator  $ {C}$ given by
\begin{equation}
 [J_3,J_+]=2\ri J_+    , \qquad  
[J_3,J_-]=-2\ri J_- ,\qquad   
[J_-,J_+]=4\ri J_3  +2\, \Xi   ,\qquad [\Xi\, , \, \cdot \,]=0,
\label{za}
\end{equation}
\begin{equation} 
 {C}=  \frac 12 (J_+ J_-  + J_-J_+ )-J_3^2 + \ri  J_3\,\Xi + \Xi^2. 
\label{zb}
\end{equation} 
Thus, $ \Xi$ is a trivial central generator such that $\mathfrak{gl}(2)\simeq \mathfrak{sl}(2,\mathbb R)\oplus \mathbb R$. As for any Lie algebra,  the coalgebra symmetry $(\mathfrak{gl}(2),\Delta)$ is defined via the primitive or non-deformed coproduct map  $\Delta$: 
\be
\Delta: \mathfrak{gl}(2)\to \mathfrak{gl}(2) \otimes \mathfrak{gl}(2) ,\qquad \Delta(X)=X \otimes 1+ 1\otimes X ,\quad X\in\{J_-,J_+,J_3,\Xi  \},
\label{zu}
\ee
which is a Lie algebra homomorphism.  Recall that the (trivial) counit and antipode can also be defined leading to a non-deformed Hopf algebra structure~\cite{CP,Abe}.

 A `one-particle' representation $D$ of $\mathfrak{gl}(2)$ (\ref{za})  in terms of the operators $(\hq_1,\hp_1)$ (\ref{b1}) reads as
 \be
 \begin{split}
 \hat J^{(1)}_-&=D(J_-)=\hq_1^2,\qquad\  \ \hat J^{(1)}_+=D(J_+)=\hp_1^2+\frac{\lambda_1}{\hq_1^2},\\
    \hat J^{(1)}_3&=D(J_3)=\hq_1\hp_1 ,\qquad \hat \Xi^{(1)}=D(\Xi)=1 .
 \end{split}
 \nonumber
 \ee
  Hence, it depends on a real parameter $\lambda_1$ which labels the representation, as the Casimir operator in  (\ref{zb}) becomes 
  \be
 \hat C^{(1)}=D(C)=\frac 12 \left(\hat J^{(1)}_+ \hat J^{(1)}_-  +\hat J^{(1)}_-  \hat J^{(1)}_+ \right) -\left(\hat J^{(1)}_3\right)^2 + \ri  \hat J_3^{(1)}\, \hat\Xi^{(1)} + \left(\hat\Xi^{(1)}\right)^2=\lambda_1.
\nonumber
\ee 
Then a `two-particle' representation $D^{(2)}=D\otimes D$ of $\mathfrak{gl}(2)$  (\ref{za}) is directly provided by the copro\-duct (\ref{zu}), namely
   \be
 \begin{split}
 \hat J^{(2)}_-&=D^{(2)}\bigl(\Delta(J_-)\bigr)=\hq_1^2+\hq_2^2,\qquad\  \ \hat J^{(2)}_+=D^{(2)}\bigl(\Delta(J_+) \bigr)=\hp_1^2+\frac{\lambda_1}{\hq_1^2}+\hp_2^2+\frac{\lambda_2}{\hq_2^2},\\
    \hat J^{(2)}_3&=D^{(2)}\bigl(\Delta(J_3)\bigr)=\hq_1\hp_1+ \hq_2\hp_2 ,\qquad \hat \Xi^{(2)}=D^{(2)}\bigl(\Delta(\Xi)\bigr)=2,
 \end{split}
 \label{zd}
 \ee
 while for the Casimir operator in (\ref{zb}), we find that
     \be
 \begin{split}
  \hat C^{(2)}&=D^{(2)}\bigl(\Delta(C)\bigr)=\frac 12 \left(\hat J^{(2)}_+ \hat J^{(2)}_-  + \hat J^{(2)}_-  \hat J^{(2)}_+ \right) -\left(\hat  J^{(2)}_3\right)^2 + \ri \hat  J_3^{(2)}\,\hat \Xi^{(2)} + \left(\hat \Xi^{(2)}\right)^2\\
  &=({\hat q_1}{\hat p_2} -
{\hat q_2}{\hat p_1})^2 + 
\otra_1\,\frac{\hat q_2^2}{\hat q_1^2}+\otra_2\,\frac{\hat q_1^2}{\hat q_2^2} + \otra_1+\otra_2+2.
 \end{split}
 \label{ze}
 \ee
 By construction~\cite{BR98,BBHMR2009}, $\hat {C}^{(2)}$  commutes with the   operators  (\ref{zd}), so that   any smooth function   ${H}$ defined on them determines  a two-dimensional  quantum {\em integrable} Hamiltonian: 
\be
\hat{H}^{(2)} =   H \left(\hat J_3^{(2)}, \hat J_+^{(2)},\hat J_-^{(2)} ,\hat \Xi^{(2)}\right).
\nonumber
\ee 
The coassociative property of the coproduct allows one to extend this result to  arbitrary dimension $d$ providing $(2d-3)$ `universal' algebraically independent operators~\cite{BBHMR2009} for $\hat{H}^{(d)}$, so that the resulting systems are called quasi-maximally superintegrable since only {\em one} additional quantum symmetry is needed to ensure maximal superintegrability. In the classical case, the central generator $\Xi$ vanishes, so that $(\mathfrak{gl}(2),\Delta)$ reduces to a Poisson $\mathfrak{sl}(2,\mathbb R)$-coalgebra, which has been deeply studied~\cite{Ballesteros2007,BBHMR2009}. Recall  also that  the Racah algebra for $\mathfrak{sl}(2,\mathbb R)$ has been widely studied in~\cite{Latini2019,Latini2021,Latini2021b}. 

In our case, let us choose ($d=2$)
\be
\hat{H}^{(2)}=  \hat J_+^{(2)} + \sum_{k=1}^N \gamma_k   \left(\hat J_3^{(2)}\right)^k  = \hat{\mathbf{p}}^2 + \sum_{k=1}^N \gamma_k ( \hat{\mathbf{q}}  \boldsymbol{\cdot} \hat{\mathbf{p}})^k  +\frac {\otra_1}{\hat q_1^2}  +\frac {\otra_2}{\hat q_2^2}.
\label{zg}
\ee 
Therefore, if we set $\lambda_1=\lambda_2=0$ we recover the generalized quantum Zernike Hamiltonian $\hcH_N  $ (\ref{b3}), while the operator $\hat C^{(2)}$ in (\ref{ze}) gives the square of the quantum angular momentum operator (\ref{b5}). Hence, for arbitrary coefficients  $\lambda_1$ and $\lambda_2$, the Hamiltonian 
 $\hat{H}^{(2)}$ is formed by the superposition of $\hcH_N$   with two Rosochatius--Winternitz potentials.  
 In the classical case with $N=2$, and within the interpretation of  $\hcH_2$ as the Hamiltonian of a curved oscillator, $\hat{H}^{(2)}$ is just the curved Smorodinsky--Winternitz system which is known to be superintegrable~\cite{Ballesteros2007};     the presence of a {\em positive} $\lambda_i$-potential corresponds to introduce a `centrifugal' barrier on 
 $ \mathbf E^2$ and $\mathbf H^2$, but a noncentral oscillator on  $\mathbf S^2$~\cite{2013Higgs}.
 From this perspective, it is rather natural to look for an  additional quantum symmetry for $\hat{H}^{(2)}$ (\ref{zg}) and to deduce and analyze the resulting spectrum by applying an algebraic approach.
 
    Finally, it would also be possible to consider a quantum deformation of $\mathfrak{gl}(2)$, inherited from a quantum $\mathfrak{sl}(2,\mathbb R)$ algebra, with quantum deformation parameter $q=\exp z$.
 In our case, it would be appropriate to make use of  the so called non-standard deformation of  $\mathfrak{sl}(2,\mathbb R)$ instead of the usual Drinfel'd--Jimbo one. The reason is that for the former the generator $J_-$ remains primitive (with trivial coproduct) and, under the representation (\ref{zd}), there appear factors of the type $\exp(z(\hq_1^2+\hq_2^2))$ explicitly in the deformed analogue of $\hat{H}^{(2)}$ (\ref{zg}); for the latter, the primitive generator is $J_3$. The quantum deformation parameter $z$ would introduce an additional perturbation from the non-deformed case $z\to 0$.
   Expected mathematical and physical properties can be guessed from already known applications of the non-standard deformation, which would be the obtainment of superintegrable Zernike-oscillator chains~\cite{Chains1999} and the construction of generalized Zernike systems on spaces of non-constant curvature determined by $z$~\cite{Chains2005}. We remark that this is still an open issue for the classical Zernike systems.

Work along these various lines is currently in progress.

 %%%%%%%%%%%%%%%%%%%%%%%%%%%%%%%%%%%%%%%%%%%%%%%%%%%%

\section*{Acknowledgements}
\phantomsection
\addcontentsline{toc}{section}{Acknowledgments}
{\small 
R.C.-S.,  F.J.H.~and  A.B.~have been partially supported by Agencia Estatal de Investigaci\'on (Spain) under  the grant PID2023-148373NB-I00 funded by MCIN/AEI/10.13039/501100011033/FEDER, UE.  F.J.H.~and A.B.~also acknowledge support  by the  Q-CAYLE Project  funded by the Regional Government of Castilla y Le\'on (Junta de Castilla y Le\'on, Spain) and by the Spanish Ministry of Science and Innovation (MCIN) through the European Union funds NextGenerationEU (PRTR C17.I1). R.C.-S. also acknowledges support  by the PID2024-156578NB-I00 funded by MICIU /AEI /10.13039/501100011033 / FEDER, EU. The research of D.L.~has been partially funded by MUR - Dipartimento di Eccellenza 2023-2027, codice CUP G43C22004580005 - codice progetto   DECC23$\_$012$\_$DIP and partially supported by INFN-CSN4 (Commissione Scientifica Nazionale 4 - Fisica Teorica), MMNLP project. D.L.~is a member of GNFM, INdAM. I.M.~has been supported by Australian Research Council Future Fellowship FT180100099. 
}

\newpage
%%%%%%%%%%%%%%%%%%%%%%%%%%%%%%%%%%%%%%%%%%%%%%%%%%

%\pagenumbering{roman}
\appendix
 \setcounter{subsection}{0}
\section{The fourth-order quantum Zernike Hamiltonian}
\label{appendixA}

 \setcounter{equation}{0}
\renewcommand{\theequation}{A.\arabic{equation}}

  We proceed similarly to   sections~\ref{s3} and \ref{s4}, first giving the quantum symmetries and the possible spectra for the quartic Hamiltonian  $\hcH_4$ in (\ref{b3}), depending on four arbitrary parameters $\gamma_1$,  $\gamma_2$, $\gamma_3$ and $\gamma_4$, and imposing that  the 
  quantum quadratic Hamiltonian  $\hcH_2$ in (\ref{b3}) is recovered for the vanishing of $\gamma_3$ and $\gamma_4$.

The two quantum symmetries (\ref{b6b}) of  $\hcH_4$, now of fourth-order in the momenta, read
\be
\begin{split}
  \hI_4 &=  \hp_2^2+ \gamma_1\hq_2\hp_2      +\gamma_2\bigl(   (\hq_1^2+\hq_2^2  ) \hp_2^2  -\hcC^2 \bigr)  \\[2pt]
   &\qquad +\gamma_3\bigl(  \hq_2^3( \hp_2^3- \hp_1^2\hp_2)+( \hq_1^3+ 3 \hq_1\hq_2^2 ) \hp_1  \hp_2^2- 3 \ri \hq_2^2\hp_2^2- 3 \ri \hq_1 \hq_2\hp_1\hp_2-\hq_2  \hp_2\bigr)     \\[2pt]
   &\qquad +\gamma_4\bigl( (\hq_2^4-\hq_1^4)( \hp_2^4-\hp_1^2 \hp_2^2  ) +4( \hq_1^3\hq_2+  \hq_1\hq_2^3)\hp_1\hp_2^3 -6\ri (\hq_2^3+ \hq_1^2\hq_2) \hp_2^3  \\[2pt]
&\qquad\qquad\qquad  -6\ri (\hq_1^3+\hq_1\hq_2^2 )\hp_1 \hp_2^2 -4(\hq_1^2+\hq_2^2)\hp_2^2  +4\hcC^2\bigr),   \\[4pt]
    \hI'_4 &= \hp_1^2+\gamma_1\hq_1\hp_1+\gamma_2\bigl(\hq_1^2+\hq_2^2 \bigr) \hp_1^2\\[2pt]
    &\qquad +\gamma_3\bigl(  \hq_1^3( \hp_1^3- \hp_1\hp_2^2)+( \hq_2^3+ 3 \hq_1^2 \hq_2)  \hp_1^2\hp_2 - 3 \ri \hq_1^2\hp_1^2- 3 \ri \hq_1 \hq_2\hp_1\hp_2-\hq_1  \hp_1\bigr) \\[2pt]
 &\qquad +\gamma_4\bigl( (\hq_1^4-\hq_2^4)( \hp_1^4-\hp_1^2 \hp_2^2 ) +4(\hq_1 \hq_2^3+ \hq_1^3 \hq_2)\hp_1^3\hp_2 -6\ri (\hq_1^3+\hq_1 \hq_2^2) \hp_1^3  \\[2pt]
&\qquad\qquad\qquad  -6\ri (\hq_2^3+\hq_1^2\hq_2 ) \hp_1^2\hp_2-4(\hq_1^2+\hq_2^2)\hp_1^2  \bigr),   
\end{split}
\nonumber
\ee
and verify the algebraic dependence  relation
$$
\hcH_4 =   \hI_4+  \hI_4' - 4 \gamma_4 \hcC^2 +\gamma_4 \hcC^4 .
$$
Note that, in this case,  the classical functions $I_4$ and   $I_4'$ 
are related to  $\mathcal I_4$ and $\mathcal I_4'$ obtained in~\cite{zk2023} through $I_4=\mathcal I_4-\gamma_2 \cC^2 +4\gamma_4 \cC^2$ and $I_4'=\mathcal I_4'$.

The operators $(\hcK_1,\hcK_2,\hcK_3)$  (\ref{b7}) fulfil the  polynomial commutation relations  given by
\be
\begin{split}
\bigl[ \hcK_1,\hcK_2 \bigr]&= \hcK_3, \qquad  \bigl[ \hcK_1,\hcK_3 \bigr]= 4   \hcK_2 -2 (\gamma_2   - 4 \gamma_4) \hcK_1^2   ,\\[2pt]
\bigl[ \hcK_2,\hcK_3 \bigr]&=    \bigl(  \gamma_1^2 +2\ri  \gamma_1\gamma_2  -4  \gamma_1\gamma_3 -8\ri   \gamma_1\gamma_4 +2 (\gamma_2+3\ri \gamma_3 -8\gamma_4)\hcH_4 \bigr) \hcK_1\\[2pt]
&\qquad+ 4(\gamma_2-4 \gamma_4) \hcK_1 \hcK_2-  2(\gamma_2 - 4 \gamma_4)\hcK_3 \\[2pt]
&\qquad + 4   \bigl( \gamma_3^2  - \gamma_1\gamma_3+\ri \gamma_2\gamma_3 - 8   \gamma_4^2  - 3\ri \gamma_1\gamma_4+2\ri \gamma_3\gamma_4  - \gamma_4 \hcH_4 \bigr)\hcK_1^3\\[2pt]
&\qquad + \bigl(   3\gamma_3^2+16\gamma_4^2 - 6 \gamma_2\gamma_4  + 6 \ri \gamma_3\gamma_4\bigr)\hcK_1^5 + 4 \gamma_4^2 \hcK_1^7.
\end{split}
\nonumber
\ee
The number and ladder operators $( \hcK,\hcK_+,\hcK_-)$ are now defined by
\be
\begin{split}
\hcK&:= \frac{1}{2 }  \hcK_1,\\
\hcK_+&:=   \hcK_2+\frac 12 \hcK_3-\biggl(  \frac{1}{2}\gamma_2-2 \gamma_4 \biggr) \hcK_1^2,\\
\hcK_-&:=  \hcK_2-\frac 12  \hcK_3- \biggl(  \frac{1}{2}\gamma_2-2 \gamma_4 \biggr)  \hcK_1^2,
\end{split}
\label{e3}
\ee
and close on  the commutation relations in (\ref{b8}) with 
 \be
 \begin{split}
  \bigl[ \hcK_-,\hcK_+ \bigr]&=2  \bigl(  \gamma_1^2 + 2 \ri \gamma_1\gamma_2 -4  \gamma_1\gamma_3-8\ri \gamma_1\gamma_4  +2 (\gamma_2+3\ri \gamma_3- 8 \gamma_4 )\hcH_4   \bigr) \hcK\\[2pt]
  &\qquad +16\bigl( \gamma_2^2+2 \gamma_3^2 -2\gamma_1\gamma_3 -6\ri \gamma_1\gamma_4 +2\ri \gamma_2\gamma_3-8\gamma_2\gamma_4 +4\ri \gamma_3\gamma_4-2\gamma_4 \hcH_4\bigr)\hcK^3 \\[2pt]
  &\qquad  + 32\bigl(  3\gamma_3^2 +16 \gamma_4^2-6  \gamma_2  \gamma_4+6\ri  \gamma_3 \gamma_4\bigr) \hcK^5+512 \gamma_4^2 \hcK^7,
 \end{split}
\nonumber
\ee
 hence  determining   a seventh-order polynomial symmetry algebra of Higgs-type   $\mathfrak{sl}^{(7)}(2,\mathbb R)$.
The structure function $\Phi$ in (\ref{b8}) turns out to be
 \be
\begin{split}
&\hcK_+\hcK_-=\Phi(\hcH_4,\hcK)\\
&=\frac{1}{4}\bigl( 4 \gamma_2 -2\ri \gamma_1  + 8\ri \gamma_3-16\gamma_4+\hcH_4\bigr)\hcH_4\\
&\qquad -   \bigl( \gamma_1^2 + 2 \ri \gamma_1\gamma_2 -4 \gamma_1\gamma_3-8\ri  \gamma_1\gamma_4  + 2(\gamma_2 +3 \ri \gamma_3 - 8 \gamma_4)\hcH_4\bigr)\hcK\\[2pt]
&\qquad  + \bigl(\gamma_1^2+4\gamma_2^2    + 2 \ri \gamma_1\gamma_2   -12\gamma_1\gamma_3 - 32 \ri \gamma_1\gamma_4+8\ri \gamma_2\gamma_3  -16 \gamma_2\gamma_4 \\[2pt]
&\qquad\qquad
 + 2(\gamma_2 +3 \ri \gamma_3 - 12 \gamma_4)\hcH_4\bigr)\hcK^2\\[2pt]
&\qquad -8\bigl( \gamma_2^2+2 \gamma_3^2-2  \gamma_1 \gamma_3-6\ri    \gamma_1 \gamma_4+2 \ri \gamma_2 \gamma_3- 8 \gamma_2 \gamma_4+ 4 \ri \gamma_3 \gamma_4- 2 \gamma_4\hcH_4 \bigr)\hcK^3\\[2pt]
&\qquad +4\bigl(  \gamma_2^2+12  \gamma_3^2+16 \gamma_4^2 -2  \gamma_1 \gamma_3-6\ri    \gamma_1 \gamma_4+2 \ri \gamma_2 \gamma_3- 28 \gamma_2 \gamma_4+ 24 \ri \gamma_3 \gamma_4-2 \gamma_4 \hcH_4\bigr) \hcK^4\\[2pt]
&\qquad -16 \bigl(  3\gamma_3^2+16\gamma_4^2-6\gamma_2\gamma_4+6\ri \gamma_3\gamma_4   \bigr)\hcK^5\\[2pt]
&\qquad
+16 \bigl(  \gamma_3^2 +24\gamma_4^2-2\gamma_2\gamma_4+2\ri \gamma_3\gamma_4 \bigr)\hcK^6 -256 \gamma_4^2  \hcK^7+64 \gamma_4^2 \hcK^8,
\end{split}
\nonumber
\ee
which further factorizes as 
\be
\begin{split}
\Phi(\hcH_4,\hcK)&=\Phi_1(\hcH_4,\hcK)\Phi_2(\hcH_4,\hcK),\qquad \bigl[\Phi_1(\hcH_4,\hcK),\Phi_2(\hcH_4,\hcK)\bigr]=0,\\
\Phi_1(\hcH_4,\hcK) &=\frac 14 \bigl( \hcH_4 - 2\ri \gamma_1 \hcK +4 \gamma_2 \hcK^2+ 8\ri \gamma_3 \hcK^3- 16 \gamma_4 \hcK^4 \bigr),\\[2pt]
\Phi_2 (\hcH_4,\hcK)&=  \hcH_4 - 2 (\ri \gamma_1-2 \gamma_2 -4\ri \gamma_3+ 8\gamma_4) {\rm Id} + 2 (\ri \gamma_1-4 \gamma_2 -12 \ri \gamma_3 +32 \gamma_4) \hcK \\[2pt]
&\qquad +4( \gamma_2+6\ri \gamma_3-24\gamma_4) \hcK^2-8(\ri \gamma_3-8\gamma_4) \hcK^3- 16\gamma_4  \hcK^4\\[2pt]
&=  \hcH_4+ 2\ri \gamma_1\bigl(\hcK  - {\rm Id} \bigr) + 4\gamma_2\bigl(\hcK  - {\rm Id} \bigr)^2-
 8\ri \gamma_3\bigl(\hcK  - {\rm Id} \bigr)^3- 16 \gamma_4\bigl(\hcK  - {\rm Id} \bigr)^4.
\end{split}
\label{e6}
\ee
The change of operators  (\ref{b10}) leads to  the deformed oscillator algebra (\ref{b11}).

 We introduce the  finite-dimensional   representation of   (\ref{b11}) into the expressions (\ref{e6}), obtaining the structure function $\Phi=\Phi_1\Phi_2$ (\ref{b12b})  as
\be
\begin{split}
\Phi_1(\B,E,u) &=\frac 14  \bigl( E - 2\ri \gamma_1 (\B+u) +4 \gamma_2  (\B+u)^2 +8\ri  \gamma_3  (\B+u)^3- 16  \gamma_4(\B+u)^4 \bigr),\\[2pt]
\Phi_2(\B,E,u) &= E + 2  \ri \gamma_1 (\B+u-1)  +4 \gamma_2   (\B+u-1)^2 -8\ri  \gamma_3  (\B+u-1)^3 -16  \gamma_4(\B+u-1)^4 .
\end{split}
\label{e7}
\ee
The spectrum $E$ of  $\hcH_4$ (\ref{b4}) is derived by imposing the  two conditions (\ref{b12}) and is established as follows.

%%%%%%%%%%%%%%%%%%%%%%%%%%%%%%%%
 
  \begin{proposition} 
\label{prop3}
(i) The set of equations (\ref{b12}) obtained from (\ref{e7}) leads to  the corresponding  solutions for the representation dependent constant $u=u(\n)$ and the spectrum $E=E(\n)$ of the Hamiltonian $\hcH_4$~(\ref{b4})  depending on the parameter  $\n\in\{1,2,\dots\}$. They provide   the structure function  $\Phi(\B,E(\n),u(\n))\equiv \Phi(\B,\n)$ for $\B=\{1,\dots,\n\}$ through (\ref{e7}).\\
(ii) Among    all the solutions, there are only two types such that the limits $\gamma_4\to 0$ and $\gamma_3\to 0$ are simultaneously well-defined for any value of $\gamma_1$ and $\gamma_2$, namely 
\be
\begin{array}{lll}
\!\!\!\!\mbox{\rm Type I:}& \displaystyle   u_{\rm I} = -\frac \n 2,&\quad \displaystyle  E_{\rm I}= -    \bigl(\ri \gamma_1 \n+\gamma_2   \n^2 - \ri \gamma_3 \n^3-   \gamma_4 \n^4 \bigr), \\[6pt]
&  \multicolumn{2}{l}{  \Phi_{\rm I}=-  \B(\B-\n-1) \biggl(\ri \gamma_1 + 2 \gamma_2   (\B-1) +\ri \gamma_3\bigr(\n(2\B-\n-2)-4(\B-1)^2 \bigl)  }\\[4pt]
&   \multicolumn{2}{l}{\qquad\qquad\qquad\qquad\qquad  +\, 4 \gamma_4 (\B-1)\bigr(\n(2\B-\n-2)-2(\B-1)^2 \bigl)\biggr)  }\\[2pt]
  &  \multicolumn{2}{l}{\qquad\qquad\qquad\qquad \times \biggl(\ri \gamma_1 - 2 \gamma_2   (\B-\n) +\ri \gamma_3\bigr(3 \n(2\B-\n)-4\B^2 \bigl)  }\\[2pt]
    &  \multicolumn{2}{l}{\qquad\qquad\qquad\qquad\qquad  -\, 4 \gamma_4 (\B-n)\bigr(\n(2\B-\n)-2\B^2 \bigl)\biggr) .}\\[8pt]
\!\!\!\! \mbox{\rm Type II:}&  \displaystyle   u_{\rm II} = -\frac \n 2,&\quad E_{\rm II}=     \ri \gamma_1 (\n +2)-\gamma_2   (\n+2)^2  - \ri \gamma_3 (\n+2)^3 +\gamma_4 (\n+2)^4,  \\[6pt]
&  \multicolumn{2}{l}{ 
\Phi_{\rm II}=- \B(\B-\n-1) \biggl(\ri \gamma_1 - 2 \gamma_2   (\B+1)  +\ri \gamma_3\bigr(\n(2\B-\n-2)-4(\B^2+\B+1) \bigl)  }\\[4pt]
&  \multicolumn{2}{l}{ \qquad\qquad\qquad\qquad\qquad 
-\, 4 \gamma_4 (\B+1)  \bigr(\n(2\B-\n-2)-2 (\B^2+1)  \bigl)\biggr) }\\[4pt]
   &  \multicolumn{2}{l}{\qquad\qquad\qquad\qquad \times   \biggl(\ri \gamma_1 + 2 \gamma_2   (\B-\n-2) +\ri \gamma_3\bigr(3(\n+2)(2\B-\n-2)-4\B^2 \bigl) } \\[2pt]
       &  \multicolumn{2}{l}{\qquad\qquad\qquad\qquad\qquad  +\,  4 \gamma_4 (\B-n-2)\bigr( (\n+2)(2\B-\n-2)-2\B^2 \bigl)\biggr) .}\\[4pt]
 \end{array}
\label{e8}   
\ee
\end{proposition}

 %%%%%%%%%%%%%%%%%%%%%%%%%%%%%%%%

The  excluded solutions exhibit divergencies, similarly to the solution (\ref{d8}) for the cubic Hamiltonian $\hcH_3$ in section~\ref{s4}, although the resulting expressions are  more complicated in this case (they involve both parameters $\gamma_3 $ and $\gamma_4$ within the denominators). In this sense, the discarded solutions in propositions~\ref{prop2} and~\ref{prop3} can be regarded as `spurious'.
 
It is worth remarking that the results given in proposition~\ref{prop3} are well-defined for {\em any} value of the four parameters $\gamma_\kk$ $(\kk=1,\dots,4)$. As expected, they  lead exactly to the results presented in proposition~\ref{prop2} for $\gamma_4=0$ and, then, to those characterized by proposition~\ref{prop1}  when    $\gamma_3=0$. In addition, it is possible to set $\gamma_3=0$ in the solutions (\ref{e8}), keeping the parameter $\gamma_4$, which determines a  superintegrable quartic perturbation of the initial Hamiltonian  $\hcH_2$.
  
  We observe that the expressions  (\ref{e8}) take real values whenever the odd coefficients $\gamma_1$ and $\gamma_3$  are pure imaginary numbers, while the even ones $\gamma_2$ and $\gamma_4$ take  real values. This difference between odd/even parameters has already been studied in the classical picture~\cite{zk2023}. Therefore, by taking into account (\ref{d9}), let us denote
\be
\gamma_1=-\rm i \beta,\qquad \gamma_2=\alpha ,\qquad \gamma_3=\ri \mmu,\qquad \gamma_4=- \nnu, \qquad \{\alpha,\beta,\mmu,\nnu\}\in\mathbb R,
\nonumber
\ee
obtaining two types of spectra for $\hcH_4$ given by
\be
\begin{split}
E_{\rm I}&= -    \bigl(\beta \n+\alpha   \n^2 +\mmu \n^3+\nnu \n^4\bigr),\\[2pt]
E_{\rm II}&=  \beta (\n +2)-\alpha   (\n+2)^2  +\mmu (\n+2)^3 -\nnu (\n+2)^4 ,
\end{split}
\nonumber
\ee
to be compared with (\ref{d10}). In addition, if we set $\alpha=-1$ and $\beta=-2$ and interpret $\n$ 
 as the principal quantum number $\fn$, we find two possible superintegrable perturbations of the spectrum (\ref{c13}) of  the Zernike Hamiltonian $\hcH_{\rm Zk}$ in (\ref{a1})   depending on two real parameters $\mmu$ and $\nnu$, and given by
    \be
E_{\rm I}=\fn(\fn+2) -\mmu \fn^3 -\nnu \fn^4,\qquad E_{\rm II}=\fn(\fn+2) +\mmu(\fn+2)^3-\nnu (\fn+2)^4.
\nonumber
     \ee

  Finally, we  remark that the results of proposition~\ref{prop3} can be applied to the isotropic oscillator
on $\mathbf E^2$, $\mathbf S^2$ and $\mathbf H^2$, described in section~\ref{s32}, thus obtaining for them different combinations of quartic/cubic perturbations  of spherical/hyperbolic type following a similar analysis as in section~\ref{s42}.

%%%%%%%%%%%%%%%%%%%%%%%%%%%%%%%%%%%%%%%%%%%%%%%%%%

 \section{The fifth-order quantum Zernike Hamiltonian}
\label{appendixB}

 \setcounter{equation}{0}
\renewcommand{\theequation}{B.\arabic{equation}}

 We now outline the main results for the case $N=5$, which are in full agreement with both conjectures proposed in section~\ref{s5}.  The quantum symmetry $\hI'_5$ in (\ref{b6b}) of $\hcH_5$ is explicitly obtained, after some cumbersome computations, as 
\be
\begin{split}
      \hI'_5 &= \hp_1^2+\gamma_1\hq_1\hp_1+\gamma_2\bigl(\hq_1^2+\hq_2^2 \bigr) \hp_1^2\\[2pt]
    &\qquad +\gamma_3\bigl(  \hq_1^3( \hp_1^3- \hp_1\hp_2^2)+( \hq_2^3+ 3 \hq_1^2 \hq_2)  \hp_1^2\hp_2 - 3 \ri \hq_1^2\hp_1^2- 3 \ri \hq_1 \hq_2\hp_1\hp_2-\hq_1  \hp_1\bigr) \\[2pt]
 &\qquad +\gamma_4\bigl( (\hq_1^4-\hq_2^4)( \hp_1^4-\hp_1^2 \hp_2^2 ) +4(\hq_1 \hq_2^3+ \hq_1^3 \hq_2)\hp_1^3\hp_2 -6\ri (\hq_1^3+\hq_1 \hq_2^2) \hp_1^3  \\[2pt]
&\qquad\qquad\quad   -6\ri (\hq_2^3+\hq_1^2\hq_2 ) \hp_1^2\hp_2-4(\hq_1^2+\hq_2^2)\hp_1^2  \bigr)
\\[2pt]
&\qquad +\gamma_5\bigl(   \hq_1^5  ( \hp_1^5+ \hp_1\hp_2^4 )- ( \hq_2^5- 5 \hq_1^4 \hq_2) ( \hp_1^4\hp_2- \hp_1^2\hp_2^3 )-(\hq_1^5 - 10 \hq_1^3 \hq_2^2- 5 \hq_1  \hq_2^4 )\hp_1^3\hp_2^2\\[2pt]
&\qquad\qquad\quad - 10 \ri \hq_1^4( \hp_1^4- \hp_1^2\hp_2^2 )- 10 \ri  (\hq_1 \hq_2^3+4 \hq_1^3 \hq_2) \hp_1^3\hp_2- 10 \ri  ( \hq_2^4+3 \hq_1^2 \hq_2^2) \hp_1^2\hp_2^2\\[2pt]
&\qquad\qquad\quad +10 \ri  \hq_1^3 \hq_2 \hp_1\hp_2^3- 25  \hq_1^3\hp_1^3- 10 (\hq_2^3+6  \hq_1^2 \hq_2)\hp_1^2\hp_2+5  (2\hq_1^3-3  \hq_1 \hq_2^2)\hp_1\hp_2^2\\[2pt]
&\qquad\qquad\quad + 15\ri \hq_1^2\hp_1^2+  15\ri \hq_1\hq_2 \hp_1\hp_2+ \hq_1\hp_1
\bigr) .
\end{split}
\nonumber 
\ee
We omit the detailed expression for $ \hI_5$, as it can be deduced from the algebraic dependence relation 
\be
\hcH_5 =   \hI_5+  \hI_5' - 4 \gamma_4 \hcC^2 +\gamma_4 \hcC^4 .
\nonumber
\ee
Next we introduce the operators $(\hcK_1,\hcK_2,\hcK_3)$ in (\ref{b7}), from which the number and ladder operators $( \hcK,\hcK_+,\hcK_-)$ with the same expression as in (\ref{e3}) can be defined. These operators close a polynomial symmetry algebra $\mathfrak{sl}^{(9)}(2,\mathbb R)$ fulfilling the commutation rules (\ref{b8}), in such a manner that the factorizing terms in (\ref{b9}) are given by
   \be
   \begin{split}
 \Phi_1  &=\frac 14 \bigl( \hcH_5 - 2\ri \gamma_1 \hcK +4 \gamma_2 \hcK^2+ 8\ri \gamma_3 \hcK^3- 16 \gamma_4 \hcK^4- 32\ri  \gamma_5 \hcK^5 \bigr),\\[2pt]
\Phi_2  &=    \hcH_5+ 2\ri \gamma_1\bigl(\hcK  - {\rm Id} \bigr) + 4\gamma_2\bigl(\hcK  - {\rm Id} \bigr)^2-
 8\ri \gamma_3\bigl(\hcK  - {\rm Id} \bigr)^3- 16 \gamma_4\bigl(\hcK  - {\rm Id} \bigr)^4+ 32\ri \gamma_5\bigl(\hcK  - {\rm Id} \bigr)^5.
\end{split}
\nonumber
\ee
From these  we directly get the relations (\ref{f2}) for $N=5$, and thus the solutions to equations (\ref{b12}). The two admissible possibilities for the spectrum that emerge imposing the constraint that they hold for any value of the coefficients $\gamma_k$ ($1\leq k\leq 5$) are
   \be
\begin{array}{lll}
 \mbox{\rm Type I:}& \displaystyle   u_{\rm I} = -\frac \n 2,&\quad \displaystyle  E_{\rm I}= -    \bigl(\ri \gamma_1 \n+\gamma_2   \n^2 - \ri \gamma_3 \n^3-   \gamma_4 \n^4+\ri \gamma_5 n^5 \bigr).\\[10pt]
  \mbox{\rm Type II:}&  \displaystyle   u_{\rm II} = -\frac \n 2,&\quad E_{\rm II}=     \ri \gamma_1 (\n +2)-\gamma_2   (\n+2)^2  - \ri \gamma_3 (\n+2)^3 +\gamma_4 (\n+2)^4+\ri \gamma_5 (n+2)^5.
  \end{array}
\nonumber
\ee
It can be directly verified that all of the  above expressions do coincide with those obtained from both conjectures in section~\ref{s5}, by entering the  value  $N=5$}.

 %%%%%%%%%%%%%%%%%%%%%%%%%%%%%%%%%%%%%%%%%%%%%%%%%%%%

\end{document}